\documentclass[10pt,journal,compsoc]{IEEEtran}
\usepackage{subfigure}
\usepackage{multirow}
\usepackage[flushleft]{threeparttable}
\usepackage{tikz}
\usepackage[normalem]{ulem}
\usepackage{booktabs}

\usepackage{amsmath}
\usepackage{amsfonts}
\usepackage{amsthm}
\usepackage{thmtools}
\usepackage{thm-restate}

\usepackage{algorithm}
\usepackage{algorithmic}

\usepackage{enumitem}
\usepackage{mathrsfs}
\usepackage{bbm}
\usepackage{hyperref}

\theoremstyle{definition}
\newtheorem{definition}{Definition}[section]
\newtheorem{prop}{Proposition}

\newcommand{\N}{\mathcal{N}} %Gaussian Distribution
\renewcommand{\P}{\mathbb{P}} %distribution
\newcommand{\Q}{\mathbb{Q}} %distribution
\DeclareMathOperator{\E}{\mathbb{E}} %expectation

\DeclareMathOperator{\MMD}{MMD}

\usepackage[font=small,labelfont=bf]{caption}

\newcommand{\pie}[1]{%
\begin{tikzpicture}
 \draw (0,0) circle (0.75ex);\fill (0.75ex,0) arc (0:(-#1+90):0.75ex) -- (0,0) -- cycle;
 \fill (0.75ex,0) arc (0:(#1-90):0.75ex) -- (0,0) -- cycle;
\end{tikzpicture}%
}

\usepackage{tcolorbox}

\begin{document}
\title{The ``Code'' of Ethics: \\A Holistic Audit of AI Code Generators}


\author{Wanlun~Ma,
        Yiliao~Song,
        Minhui~Xue,
        Sheng~Wen,
    and~Yang~Xiang,~\IEEEmembership{Fellow,~IEEE}
\thanks{W. Ma, S. Wen and Y. Xiang are with the School of Science, Computing and Engineering Technologies, Swinburne University of Technology, VIC 3122 Australia.
E-mail:\{wma, swen, yxiang\}@swin.edu.au.}%\protect\\
\thanks{Y. Song is with the College of Business and Law, RMIT University, VIC 3001, Australia.
E-mail: yiliao.song@rmit.edu.au} %\protect\\
\thanks{M. Xue is with the Cybersecurity and Quantum Systems Group, CSIRO's Data61. 
E-mail: minhuixue@gmail.com}%\protect\\
\thanks{Corresponding Author: S. Wen}%\protect\\
\thanks{Manuscript received April 19, 2021; revised August 16, 2021.}}

\markboth{Journal of \LaTeX\ Class Files,~Vol.~14, No.~8, August~2021}%
{Shell \MakeLowercase{\textit{et al.}}: A Sample Article Using IEEEtran.cls for IEEE Journals}

\IEEEpubid{0000--0000/00\$00.00~\copyright~2021 IEEE}

\IEEEtitleabstractindextext{%
\begin{abstract}
AI-powered programming language generation (PLG) models have gained increasing attention due to their ability to generate source code of programs in a few seconds with a plain program description. Despite their remarkable performance, many concerns are raised over the potential risks of their development and deployment, such as legal issues of copyright infringement induced by training usage of licensed code, and malicious consequences due to the unregulated use of these models. In this paper, we  present the first-of-its-kind study to systematically investigate the accountability of PLG models from the perspectives of both model development and deployment. In particular, we develop a holistic framework not only to audit the training data usage of PLG models, but also to identify neural code generated by PLG models as well as determine its attribution to a source model. To this end, we propose using membership inference to audit whether a code snippet used is in the PLG model's training data. In addition, we propose a learning-based method to distinguish between human-written code and neural code. In neural code attribution, through both empirical and theoretical analysis, we show that it is impossible to reliably attribute the generation of one code snippet to one model. We then propose two feasible alternative methods: one is to attribute one neural code snippet to one of the candidate PLG models, and the other is to verify whether a set of neural code snippets can be attributed to a given PLG model. The proposed framework thoroughly examines the accountability of PLG models which are verified by extensive experiments. The implementations of our proposed framework are also encapsulated into a new artifact, named \textsc{CodeForensic}, to foster further research.

\end{abstract}

\begin{IEEEkeywords}
AI Code Generator, Programming Language Generation, Accountability
\end{IEEEkeywords}}

\maketitle

\section{Introduction}

Recent advancements in deep learning have powered the development of Programming Language Generation (PLG) models, also known as AI code generators, to support code intelligence. Examples of these models include aiXcoder~\cite{aixcoder}, GitHub Copilot~\cite{copilot}, and ChatGPT~\cite{OpenAI2022ChatGPT}.
Trained on vast quantities of code datasets collected from open-source repositories, PLG models can generate a program code in response to a code prompt, such as the description of a programming problem or a code context immediately. 
Despite their remarkable performance, many have raised concerns over the potential risk of their development and deployment~\cite{henderson2023foundation}.

\begin{figure}[tp]
\centering
\includegraphics[width=1\linewidth]{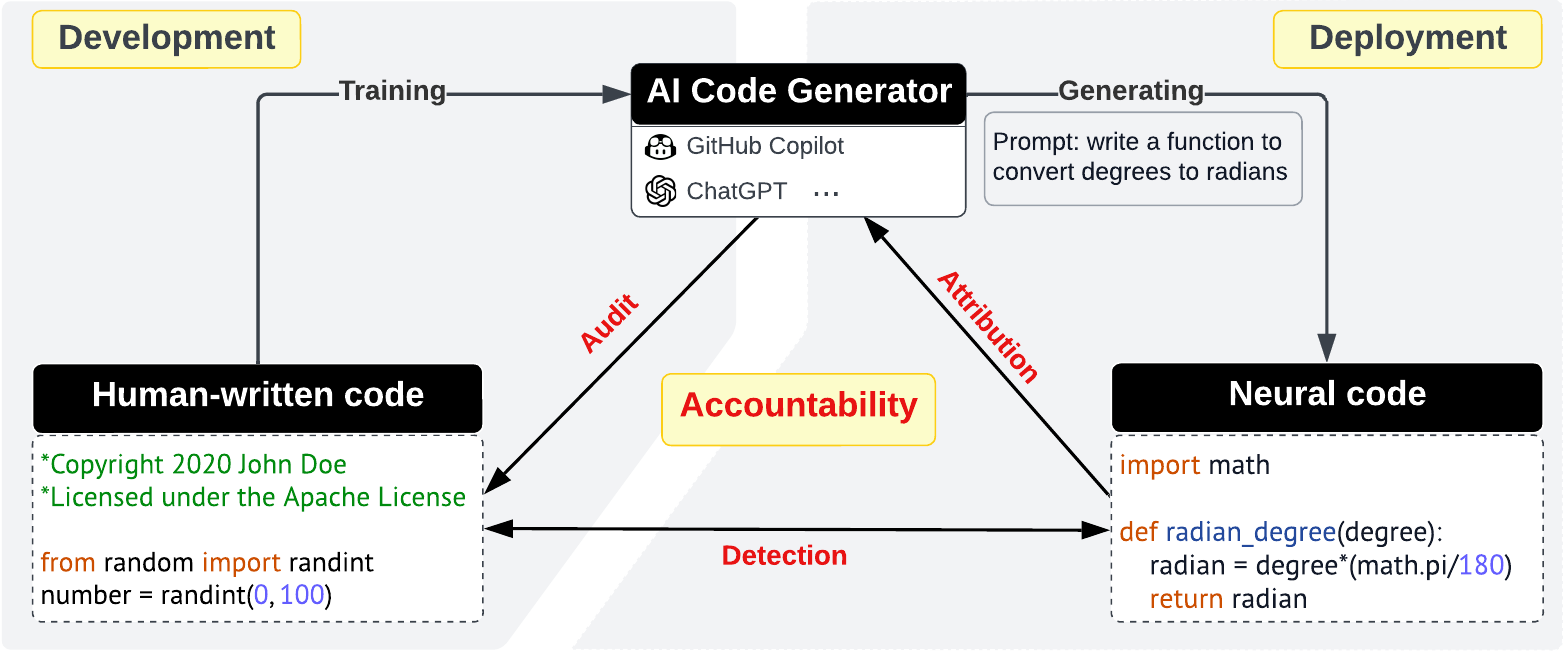}
% \vspace{-1mm}
\caption{\label{fig: overview} An overview of our study on accountability of AI code generators.}
\vspace{-3mm}
\end{figure}

From the aspect of model development, unauthorized use of source code for training PLG models can lead to copyright infringement and potentially harm the intellectual property of code creators.
PLG models are capable of generating code that is identical to licensed source code but without the corresponding attribution to the original source coder~\cite{henderson2023foundation}. Moreover, source code used in PLG models is often subject to strict licenses that impose regulations such as commercial usage, modification, and source attribution requirements. Unauthorized usage of source code in PLG models can raise legal and ethical issues, especially if the generated content is not transformative (\textit{i.e.}, the model outputs similar content to copyrighted training data)~\cite{henderson2023foundation}. For example, GitHub Copilot has been found to generate large segments of copied code without attribution to licensed source code~\cite{Ronacher2021twitter, Davis2022twitter}, leading to a lawsuit against it for alleged copyright infringement~\cite{DOEvGitHub}. OpenAI, the owner of ChatGPT, is also facing potential fines and a ban in Europe if it fails to convince the authorities its data use practices are legal~\cite{Heikkila2023OpenAI}.
To mitigate this infringement harm, the U.S. Copyright Office recently launched an initiative to examine copyright law and policy issues over the use of copyrighted materials in generative AI training~\cite{copyrightoffice2023}. 
We notice that copyright protection for code is more complex and challenging than that for creative works such as text or music, which have had extensive legal and technical measures in place to protect them~\cite{samuelson2017saving, henderson2023foundation}. 
Although text and code generation have some similarities in how their models are trained, they have each spawned distinctive case law with different assessments when it comes to fair use of copyrighted material~\cite{henderson2023foundation}.
As a result, it is difficult for any arbitration committee or court to apply standard copyright principles, such as idea and/or expression distinctions in text- and music-related cases, into code-related cases due to the technical sophistication and functionality of software program~\cite{bloch2022some}.
Therefore, it remains a complex and open challenge for collecting digital forensics to detect and prevent copyright violations in \textit{code domain}.

The potential misuse of PLG models also raises ethical concerns that require attention, particularly for model deployment~\cite{bommasani2021opportunities}. One of the major issues is the generation of false information~\cite{borji2023categorical, shen2023chatgpt} and artificial hallucinations~\cite{alkaissi2023artificial}, which can lead to the spread of misinformation. Not only does this have the potential to harm the public interest, but it can also disrupt the network environment~\cite{chaoning2022one}. For instance, to avoid being overwhelmed by the plausible but incorrect AI-generated code, Stack Overflow, a programming-related question-and-answer (Q$\&$A) website recently banned ChatGPT-generated responses~\cite{stackoverflow2022Temporary}. In addition, PLG models can be used for academic plagiarism. Recent research~\cite{biderman2022fooling} has demonstrated that large PLG models such as GPT-J~\cite{wang2021gpt} can accomplish computer science assignments for students without triggering MOSS~\cite{schleimer2003winnowing}, a widely used academic plagiarism detection tool. To address the potential misuse of PLG models and ensure their responsible use, it is necessary to detect the neural code generated by PLG models and attribute it to the source model.

There is a tangible demand from the regulatory authorities and software industry to develop a holistic framework to examine the accountability of PLG models in terms of auditing training data used by PLG models, detecting neural code generated by PLG models, and attributing it to its source model.
Despite much effort made in this area, they solely focus on natural language generation models~\cite{gehrmann2019gltr, zellers2019defending, ippolito2020automatic, uchendu2020authorship, munir2021through, hisamoto2020membership, song2019auditing}. Less attention has been paid to emerging PLG models. Additionally, existing studies always investigate model accountability from a single aspect, either the training data~\cite{hisamoto2020membership, song2019auditing} or the model output~\cite{gehrmann2019gltr, zellers2019defending, ippolito2020automatic, uchendu2020authorship, munir2021through}. This prompts the need to examine the accountability of PLG models from a holistic perspective.

\noindent \textbf{Our Work.} In this paper, we systematically study the accountability of PLG models from the perspectives of both model development and deployment. In particular, our goal is not only to audit the training data usage of PLG models, but also to detect neural code as well as determine its attribution to a source model. Therefore, as shown in Figure~\ref{fig: overview}, we seek to investigate the following research questions (\textbf{RQ}s):
\begin{itemize} %[leftmargin=*]
\item \textbf{RQ1: Training Usage Audit.} Given a code snippet, can we determine whether this code snippet is used in the PLG model’s training data?
\item \textbf{RQ2: Neural Code Detection.} Given a code snippet, can we determine whether this code snippet is written by a human or generated by some model?
\item \textbf{RQ3: Neural Code Attribution.} Given a neural code snippet and a PLG model, can we determine whether the code snippet is generated by the model?
\end{itemize}

For \textit{training usage audit} (\textbf{RQ1}), we perform a membership inference based on likelihood ratio test (LRT) to determine whether or not a code snippet exists in a model’s training dataset. Experimental results show that the LRT-based membership inference approach can achieve a good performance. 
Moreover, considering the limitations of current membership inference methods, we provide several concrete recommendations from both technical and regulatory perspectives to enhance transparency and accountability of PLG models.

For \textit{neural code detection} (\textbf{RQ2}), We propose a learning-based classifier to distinguish between human-written code and neural code. Our extensive evaluations on five state-of-the-art PLG models and three benchmark datasets demonstrate the existence of common fingerprints shared across different PLG models. 
This finding can help software communities detect the neural code generated by PLG models and mitigate their threat to the credibility of Q$\&$A websites. 
Moreover, our experimental result shows that even a well-performing neural text detector cannot effectively distinguish the neural code generated by PLG models, highlighting the critical need to investigate the features of PLG models and their generated neural code.

We then delve deep to investigate \textit{neural code attribution} (\textbf{RQ3}). Here, we aim to determine the attribution of a neural code snippet to a certain suspected model. 
We not only empirically demonstrate that it is impossible to reliably solve this one-to-one attribution problem (that is, \emph{it is impossible to reliably attribute ONLY one neural code snippet to a given PLG model}), but also provide pioneering theoretical support to this claim. 

Hence, we consider two feasible situations where we relax the constraint on either the number of PLG models or the number of neural code snippets.
By relaxing the constraint on the number of PLG models, we propose to address \textit{attribution classification} by using a learning multi-class classifier to predict the source model (among a set of candidate models) for a given neural code snippet. 
Our empirical results demonstrate that different PLG models have distinctive fingerprints which are inherited by their corresponding generated neural code. 
This allows us to attribute the neural code snippet to one in a set of candidate PLG models.
On the other hand, by relaxing the constraint on the number of neural code snippets, we address \textit{attribution verification} by formulating it as a two-sample hypothesis testing problem. Specifically, we employ the Maximum Mean Discrepancy ($\MMD$) test to verify whether a set of neural code snippets can be attributed to a given PLG model. We theoretically prove that our $\MMD$ test can approach a 100\% success rate when we have sufficient neural code snippets (around 30, as suggested by our empirical studies).

\begin{table*}[htp]
\centering
\caption{\label{table: summary of code generation models} A summary of the PLG models studied in our work.}
\vspace{-1mm}
\resizebox{0.95\textwidth}{!}{
\begin{threeparttable}
\begin{tabular}{lllllllllllll}
\toprule
\multicolumn{1}{c}{\textbf{Architecture}} & \textbf{Algorithm} & \textbf{ckpt-name} & \textbf{Training dataset} & \textbf{python} & \textbf{java} & \textbf{js} & \textbf{php} & \textbf{ruby} & \textbf{go} & \textbf{c} & \textbf{c\#} & \textbf{c++} \\
\midrule
\multirow{6}{*}{\textbf{decoder-only}} & \multirow{3}{*}{\begin{tabular}[c]{@{}l@{}}CodeGen\\ (350M/2B/6B/16B)\end{tabular}} & CodeGen-NL & ThePile (825G) &  \pie{45} & \pie{45} & \pie{45} & \pie{45} & \pie{45} & \pie{45} & \pie{45} & \pie{45} & \pie{45} \\
 &  & CodeGen-Multi & BigQuery & \pie{180} & \pie{180} & \pie{180} & \pie{45} & \pie{45} & \pie{180} & \pie{180} & \pie{45} & \pie{180} \\
 &  & CodeGen-Mono & BigPython & \pie{360} & \pie{180} & \pie{180} & \pie{45} & \pie{45} & \pie{180} & \pie{180} & \pie{45} & \pie{180} \\
\cmidrule{2-13}
 & InCoder & InCoder-1B/6B & 216GB & \pie{180} & \pie{180} & \pie{180} & \pie{180} & \pie{45} & \pie{180} & \pie{180} & \pie{45} & \pie{180} \\
\cmidrule{2-13}
 & PolyCoder & PolyCoder-160M/0.4B/2.7B & 249GB & \pie{180} & \pie{180} & \pie{180} & \pie{180} & \pie{180} & \pie{180} & \pie{180} & \pie{180} & \pie{180} \\
\cmidrule{2-13}
 & GPT-Neo & GPT-Neo-125M/1.3B/2.7B & ThePile & \pie{45} & \pie{45} & \pie{45} & \pie{45} & \pie{45} & \pie{45} & \pie{45} & \pie{45} & \pie{45} \\
\midrule
\multicolumn{1}{c}{\multirow{1}{*}{\textbf{encoder-decoder}}} & \multirow{1}{*}{CodeT5} & CodeT5-small/base & CNS+C/C\# & \pie{180} & \pie{180} & \pie{180} & \pie{180} & \pie{180} & \pie{180} & \pie{180} & \pie{180} & \pie{90} \\
\bottomrule
\end{tabular}
\begin{tablenotes}
\item[]\footnotesize\pie{90}: the language is not contained in the training dataset; \pie{45}: the proportion of the language is small (less than 2\%);\pie{180}: the proportion of the language is about 2\% to 50\%; \pie{360}: the proportion of the language is lager than 50\%.
\end{tablenotes}
\end{threeparttable}
}
\vspace{-2mm}
\end{table*}

\noindent \textbf{Contributions.} 
Our main contributions are outlined below:
\begin{itemize} %[leftmargin=*]
\item To the best of our knowledge, we present the first-of-its-kind study to systematically investigate the accountability of PLG models from both model development and deployment perspectives.

\item We pioneer conducting the first study on the copyright infringement issue of current PLG models, and propose an LRT-based membership inference to audit training usage of PLG models.

\item For neural code detection, we propose a learning-based method by a classifier that can distinguish between human-written code and neural code. Extensive evaluations show the common fingerprints shared among different PLG models, allowing us to effectively detect neural code.

\item For neural code attribution, we demonstrate, empirically and theoretically, that it is impossible to reliably attribute the generation of one code snippet to a specific model, shedding light on the challenges of dissecting neural code accountability.

\item We propose two feasible alternative methods for dissecting neural code attribution: \textit{attribution classification} and \textit{attribution verification}. Extensive evaluations demonstrate the effectiveness and robustness of the proposed methods.

\item We make our tool and methods publicly available, under the name of \textsc{CodeForensic},\footnote{\url{https://github.com/codeforensic/CodeForensic}} which is a toolkit to characterize neural code accountability.

\end{itemize}

\section{Preliminaries}

\subsection{Programming Language Generation Models}

In this work, we focus on PLG models that synthesize code from natural language descriptions of programming problems or code context. In this paper, we consider five state-of-the-art code generation models that are publicly available in the open-source community. Details of these models are summarized in Table~\ref{table: summary of code generation models}.

{$\vcenter{\hbox{\tiny$\bullet$}}$} \textbf{InCoder} \cite{fried2022incoder} Facebook's InCoder is a generative model for program synthesis and editing, using the same architecture as the dense models described in \cite{artetxe2021efficient}. Its training data consists of online open-source code from GitHub and GitLab (28 languages in total) with the great majority of files being Python and JavaScript.

{$\vcenter{\hbox{\tiny$\bullet$}}$} \textbf{GPT-Neo} \cite{black2021gpt} GPT-Neo is a GPT-3-style model based on the EleutherAI's replication. It is trained on ThePile dataset \cite{gao2020pile}, a large corpus containing a blend of natural language texts and code from various domains.

{$\vcenter{\hbox{\tiny$\bullet$}}$} \textbf{CodeGen} \cite{nijkamp2022codegen} CodeGen is a standard transformer-based autoregressive model for program synthesis released by Salesforce. CodeGen is trained sequentially on three datasets (\textit{i.e.}, ThePile \cite{gao2020pile}, BigQuery \cite{bigquery}, and BigPython \cite{nijkamp2022codegen}), resulting in three corresponding models CodeGen-NL, CodeGen-Multi, and CodeGen-Mono, respectively.

{$\vcenter{\hbox{\tiny$\bullet$}}$} \textbf{PolyCoder} \cite{xu2022systematic} PolyCoder builds on the GPT-2 architecture \cite{radford2019language} and is trained on a code dataset of 249 GB, which is collected from GitHub in 12  programming languages.

{$\vcenter{\hbox{\tiny$\bullet$}}$} \textbf{CodeT5} \cite{wang2021codet5} Different from the above encoder-only code generation models, Salesforce's CodeT5 is an encoder-decoder transformer-based model with the same architecture as T5~\cite{raffel2020exploring}. It is trained on CodeSearchNet~\cite{husain2019codesearchnet} and two collected datasets of C/C\# from BigQuery~\cite{bigquery}. 

We note that our goal is to dissect the accountability of PLG models in multiple programming languages. Therefore, we do not include models that are trained solely on code data in Python or Java languages, such as CodeParrot~\cite{tunstall2022natural}, CodeGPT~\cite{lu2021codexglue}, PyCodeGPT~\cite{Zan2022CERT},
PLBART~\cite{ahmad2021unified}, and \textsc{PyMT5}~\cite{clement2020pymt5}.

\subsection{Datasets}

We conduct our experiments on the following code generation benchmark datasets: HumanEval \cite{chen2021evaluating}, MBPP \cite{austin2021program}, APPS \cite{hendrycks2021apps}, and MBXP \cite{athiwaratkun2022multi}. These benchmark datasets are commonly involved in prior code-generation-related studies \cite{fried2022incoder, nijkamp2022codegen, xu2022systematic, wang2021codet5}.

{$\vcenter{\hbox{\tiny$\bullet$}}$} \textbf{HumanEval} \cite{chen2021evaluating} HumanEval dataset released by OpenAI consists of 164 programming problems with hand-written solutions in Python language.

{$\vcenter{\hbox{\tiny$\bullet$}}$} \textbf{MBPP} \cite{austin2021program} 
The Mostly Basic Programming Problems (MBPP) dataset released by Google includes 974 crowd-sourced programming problems with natural language descriptions in English and corresponding hand-written solutions in Python.

{$\vcenter{\hbox{\tiny$\bullet$}}$} \textbf{APPS} \cite{hendrycks2021apps} The Automated Programming Progress Standard (APPS) dataset consists of 10,000 coding problems of varying difficulty levels. These coding problems are collected from different open-access coding websites.

{$\vcenter{\hbox{\tiny$\bullet$}}$} \textbf{MBXP} \cite{athiwaratkun2022multi} Different from the above three Python-specific datasets, Most Basic X(Java/Go/Ruby, etc.) Programming Problems (MBXP) is a collection of multi-lingual datasets. Each sub-dataset (\textit{e.g.}, MBXP-Java) is generated by a conversion framework that translates prompts and test cases from the original Python dataset MBPP to the corresponding data in a target language.

\section{Training Usage Audit}
The powerful memorization of PLG models poses a risk of reproducing data directly from the training data, which potentially raises legal concerns over copyright infringement and ownership~\cite{henderson2023foundation, chen2023pathway}. In this section, we investigate the legal and ethical issues of current PLG models from the perspective of model development which is to audit the training data usage of PLG models. We also propose to use the membership inference approach to answer \textbf{RQ1}: Is a given code snippet used in the PLG model’s training data?

\subsection{Intuition and Goals}
\label{sec: Training Usage Auditing Intuition and Goals}
Membership inference utilizes the difference in the model's behaviors on training data (labeled as ``member'') and others (labeled as ``non-member'') ~\cite{shokri2017membership}, as the models are inclined to behave more confidently on member data than on non-member data. 
In other words, the model's loss on an individual data sample directly reflects how well the model has memorized this data sample.
Therefore, we can utilize the membership inference to audit the training usage of the PLG model, where we aim to achieve the following goal:

{$\vcenter{\hbox{\tiny$\bullet$}}$} 
\textbf{Determining Membership.} The goal is to determine whether a given (human-written) code snippet is used for training the PLG model.

\subsection{Methodology}
\label{sec: Training Usage Auditing Methodology}

Model loss-based membership inference methods predict the data sample with a lower loss to be a member of the training data~\cite{jagannatha2021membership, yeom2018privacy}. 
Although the model’s loss can reflect the membership status of each individual target data, recent research~\cite{mireshghallah2022quantifying, ye2022enhanced, carlini2022membership} show that prior methods that rely solely on the loss of the target model yield sub-optimal performance since the loss value provides a weak indicator for membership prediction.
Therefore, we utilize an advanced reference-model-based likelihood ratio test~\cite{ye2022enhanced, murakonda2021quantifying}. According to the Neyman-Pearson lemma~\cite{neyman1933ix}, membership inference based on likelihood ratio test can obtain the highest testing power (\textit{i.e.}, achieve optimal performance)~\cite{carlini2022membership}. Specifically, We use the likelihood ratio test to distinguish the following two hypotheses:

{$\vcenter{\hbox{\tiny$\bullet$}}$} 
Null hypothesis ($H_{out}$): The code snippet $x$ is drawn from the general population, independently from the training dataset of the PLG model.

{$\vcenter{\hbox{\tiny$\bullet$}}$} 
Alternative hypothesis ($H_{in}$): The code snippet $x$ is drawn from the training dataset of the target PLG model.

Let $\mathbf{G}_T$ and $\mathbf{G}_R$ be the target and reference PLG model, respectively. The reference model $\mathbf{G}_R$ is trained on data samples drawn from the general population. The likelihoods for $H_{out}$ and $H_{in}$ are denoted as $Pr[x; \mathbf{G}_R]$ and $Pr[x; \mathbf{G}_T]$, respectively. Therefore, the log likelihood statistic is computed as follows
\begin{align}
& L(x) = \log \left(\frac{Pr[x; \mathbf{G}_R]}{Pr[x; \mathbf{G}_T]}\right)
\end{align} 
By performing the likelihood ratio test, we can decide whether or not to reject the null hypothesis $H_{out}$. 

\noindent \textbf{Computing Likelihood.} Generally, the PLG model is also a language model (LM). For a sequence of $t$ discrete tokens $x$=$\left\{w_1, \ldots, w_1\right\}$, LMs can assign it a probability, which measures how confident a language model predicts such a sequence of tokens~\cite{bengio2000neural}. The probability $P(w_1,...,w_t)$ is computed by applying the chain rule of probability:
\begin{align}
\label{equ: language model}
    P(w_1, \ldots, w_t) = \prod^{t}_{i=1}P(w_i|w_1, \ldots, w_{i-1})
\end{align}
Rather than directly computing Eq.~(\ref{equ: language model}), a more commonly used metric is $Perplexity$ metric~\cite{manning1999foundations, huggingface2022Perplexity}. A lower perplexity indicates higher model confidence, \textit{i.e.}, higher likelihood of the code snippet. The $Perplexity$ metric is computed as:
\begin{align}
\label{equ: perplexity}
    PPL(x) = \exp\left\{-\frac{1}{t}\sum^{t}_{i=1}\log P(w_i|w_1, \ldots, w_{i-1})\right\}
\end{align}

Hence, we calculate the log likelihood statistic using the perplexity value:
\begin{align}
L(x) &= \log \left(\frac{Pr[x; \mathbf{G}_R]}{Pr[x; \mathbf{G}_T]}\right) \\
&= \log \left(\frac{PPL^{-1}(x; \mathbf{G}_R)}{PPL^{-1}(x; \mathbf{G}_T)}\right)
\end{align}

\subsection{Experiment Setup}

\noindent \textbf{Dataset Configurations.} 
We use the test split of the MBPP dataset (500 coding problems) as non-member data, while the remainder as member data (474 coding problems). We also use the APPS dataset which contains a train and test splits with 5000 samples each.
We do not directly use the original training dataset for membership inference as none of the PLG models in Table~\ref{table: summary of code generation models} release their training data.

\noindent \textbf{Model Configurations.}. We choose CodeGen, PolyCoder, and GPT-Neo as the target models, while we use InCoder as the domain-specific reference model. We also use the standard pre-trained GPT-2 as a general-domain reference model for ablation study. We fine-tune the target models on the training dataset using Adam optimizer with a learning rate of $10^{-5}$.

\begin{table}[]
\centering
\caption{\label{tabel: detection_results_q4} The AUC scores of different membership inference approaches, including model loss-based (LOSS-Based), likelihood ratio test using a general-domain reference model (LRT-G), and the one using a domain-specific model (LRT-S).}
\vspace{-1mm}
\begin{threeparttable}
\begin{tabular}{ccccc}
\toprule
 Dataset $\downarrow$ & Target Model $\rightarrow$ & CodeGen & PolyCoder & GPT-Neo \\ 
 \midrule
\multirow{3}{*}{APPS} & LOSS-Based & 55.9 & 62.5 & 60.1 \\ 
 & LRT-G & 73.2 & 78.1 & 83.6 \\  
 & LRT-S & \textbf{79.2} & \textbf{85.3} & \textbf{86.1} \\ 
\midrule
\multirow{3}{*}{MBPP} & LOSS-Based & 69.1 & 61.9 & 67.8 \\ 
 & LRT-G & 80.0 & 72.8 & 78.8 \\ 
 & LRT-S & \textbf{83.3} & \textbf{76.2} & \textbf{86.8} \\ 
\bottomrule
\end{tabular}
\end{threeparttable}
\vspace{-2mm}
\end{table}

\subsection{Results \& Analysis}
\label{Training Usage Auditing Results and Analysis}

This section presents the main results of the membership inference. Table~\ref{tabel: detection_results_q4} shows the performance of different membership inference approaches, including target model loss-based method (LOSS-Based)~\cite{jagannatha2021membership}, likelihood ratio test using a general-domain reference model (LRT-G), and likelihood ratio test using a domain-specific reference mode (LRT-S). 
We can observe that LRT-based methods (LRT-S and LRT-G) significantly outperform the LOSS-based one. For example, the AUC scores of LRT-S and LRT-G are at least 10\% higher than those of the LOSS-Based method on all datasets and target models. 
This indicates that LRT methods can better differentiate similar samples than loss-based methods because the reference model used in LRT methods can magnify the difference of the target model's behavior between members and non-members.

In addition, we find that it is better to use a code-domain reference model when using the LRT-based membership inference to answer RQ1.
As we can see in Table~\ref{tabel: detection_results_q4}, LRT-G using the GPT-2 model achieves comparable performance, but is worse than LRT-S using the InCoder model. This is mostly due to the overlap in domains between the InCoder reference model and the target models. All of these models are trained on the programming language domain except that GPT-2 is trained on the natural language domain. This indicates that an ideal reference model for this approach would be trained on a similar data distribution to that of the target model in order to better characterize the data memorization of the target model. 

\subsection{Discussion}
\label{sec: Training Usage Auditing Discussion}

Both the LOSS-based method and LRT-based method are considered as black-box membership inference in previous research~\cite{jagannatha2021membership, ye2022enhanced}, where they assume only access to the probability value from the target model but not to its parameters or internal computations. 
However, it is not practical to obtain probability value from the PLG-model-based services, since users can only obtain the predicted results (generated code) from such black-box services.
Although existing shadow-model-based membership inference can perform audits using only the generated content, they either yield poor performance on large transformer-based language models~\cite{hisamoto2020membership} or merely focus on shallow and simple LSTM models~\cite{song2019auditing}.

To ensure regulatory compliance of PLG models with respect to training data usage, we recommend concerted efforts from both technical and regulatory perspectives. First, there is a pressing need to carry out more research on developing robust membership inference methods that can effectively audit PLG models using only generated content from PLG models. Second, given the limitations of current membership inference methods, regulators can mandate PLG model developers to provide internal access to their models to allow for a reliable audit of the training data usage. Third, model developers can be required to publish or provide clear documentation on the training data used for PLG models, including the sources and legal status of the data. These actions will enhance transparency and accountability in the development and deployment of PLG models, and help mitigate potential legal and ethical risks.

\begin{tcolorbox}
\begin{itemize} [leftmargin=*]
\item Given access to the probability value, the state-of-the-art LRT-based membership inference method performs well on auditing training data usage of PLG models. 
\item Limitations of current membership inference methods in practical application emphasize the need for concerted efforts from technical and regulatory perspectives to enhance accountability in the development of PLG models.
\end{itemize}
\end{tcolorbox}

\section{Neural Code Detection}
\label{sec: Neural Code Detection}

Aside from the accountability of model development, the potential misuse of PLG models motivates us to investigate their accountability in terms of model deployment. To this end, our study of model deployment accountability starts with Neural Code Detection (\textbf{RQ2}) that questions how to distinguish between neural code and human-written code. 
To answer this question, we propose a learning-based method that builds a detector to classify a code snippet as human-written or model-generated.
Specifically, this section begins with our intuitions and goals, followed by the pipeline of building the detector, and finally presents the experimental setup and the evaluation results.

\subsection{Intuitions and Goals}
\label{sec: neural code detection Intuition and Goals}

Our intuition behind the neural code detection comes from research in natural language generation (NLG) models \cite{zellers2019defending, uchendu2020authorship, ippolito2020automatic, rodriguez2022cross}, and specifically driven by their conclusion that different NLG models share common or similar fingerprints.
We infer that PLG models have similar behaviors in this aspect.
Namely, PLG models share common or similar fingerprints, and these fingerprints can be leveraged to detect neural code from human-written code.
Therefore, our learning-based method for neural code detection aims to achieve the following two goals:

{$\vcenter{\hbox{\tiny$\bullet$}}$} \textbf{Detect neural code generated by known models.} The primary goal for detection is to distinguish neural code from human-written code where the neural code for training and testing is generated by the same PLG model. 

{$\vcenter{\hbox{\tiny$\bullet$}}$} \textbf{Detect neural code generated by unseen models.} Due to the growing proliferation of PLG models, it is impossible to include every PLG model when learning the detection model. Therefore, the detection model should be able to detect neural code generated by unseen models. 
This can verify that the common fingerprints captured by the detector generally exist in all PLG models rather than in some specific models.

\begin{figure*}[tp]
\centering
\subfigure[HumanEval]{
\begin{minipage}[]{0.32\textwidth}
\centering
\includegraphics[width=0.96\textwidth]{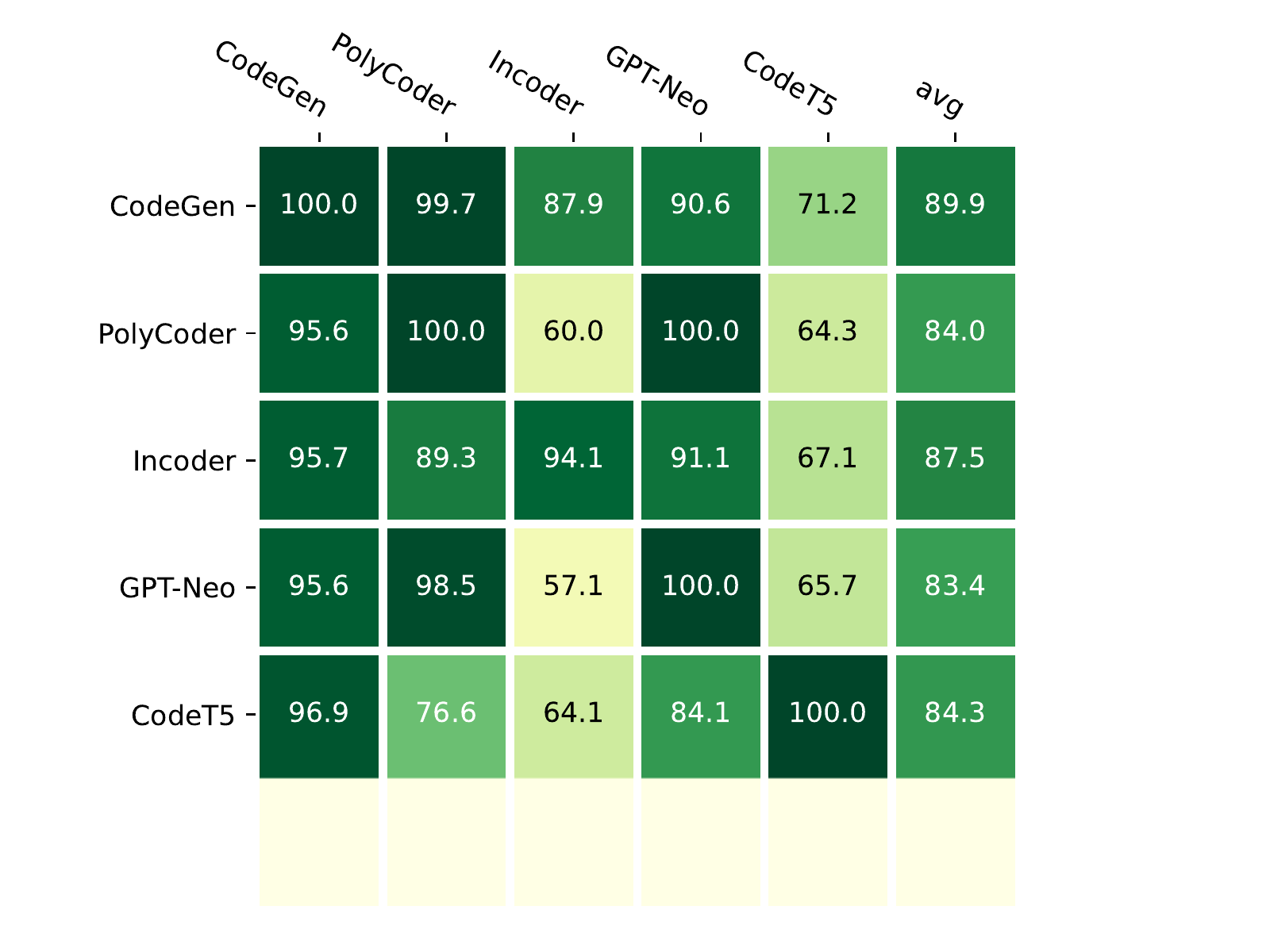}
\label{fig: detection_results_q1_humaneval}
\end{minipage}}
\subfigure[MBPP]{
\begin{minipage}[]{0.32\textwidth}
\centering
\includegraphics[width=0.96\textwidth]{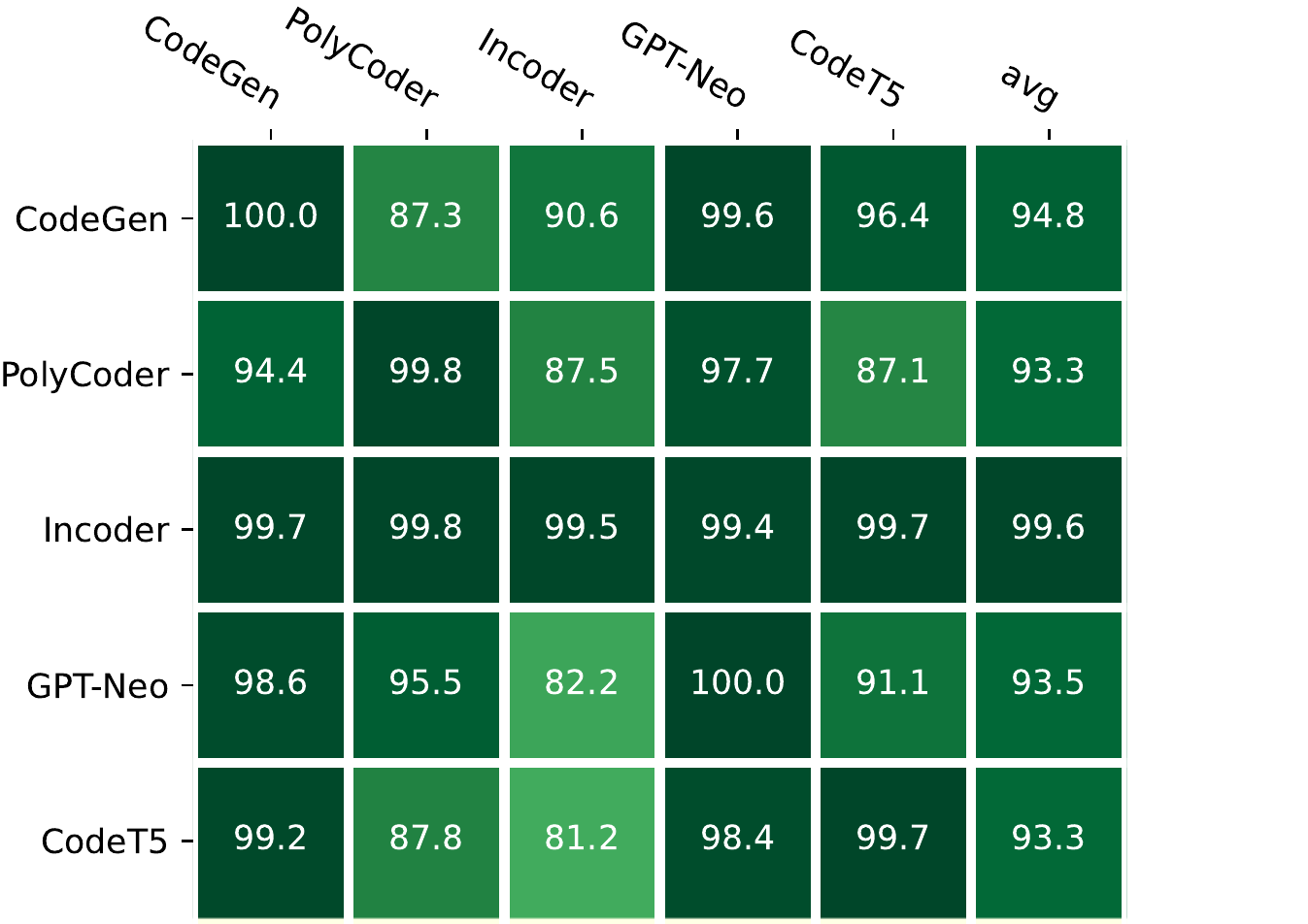}
\label{fig: detection_results_q1_mbpp}
\end{minipage}}
\subfigure[APPS]{
\begin{minipage}[]{0.32\textwidth}
\centering
\includegraphics[width=0.96\textwidth]{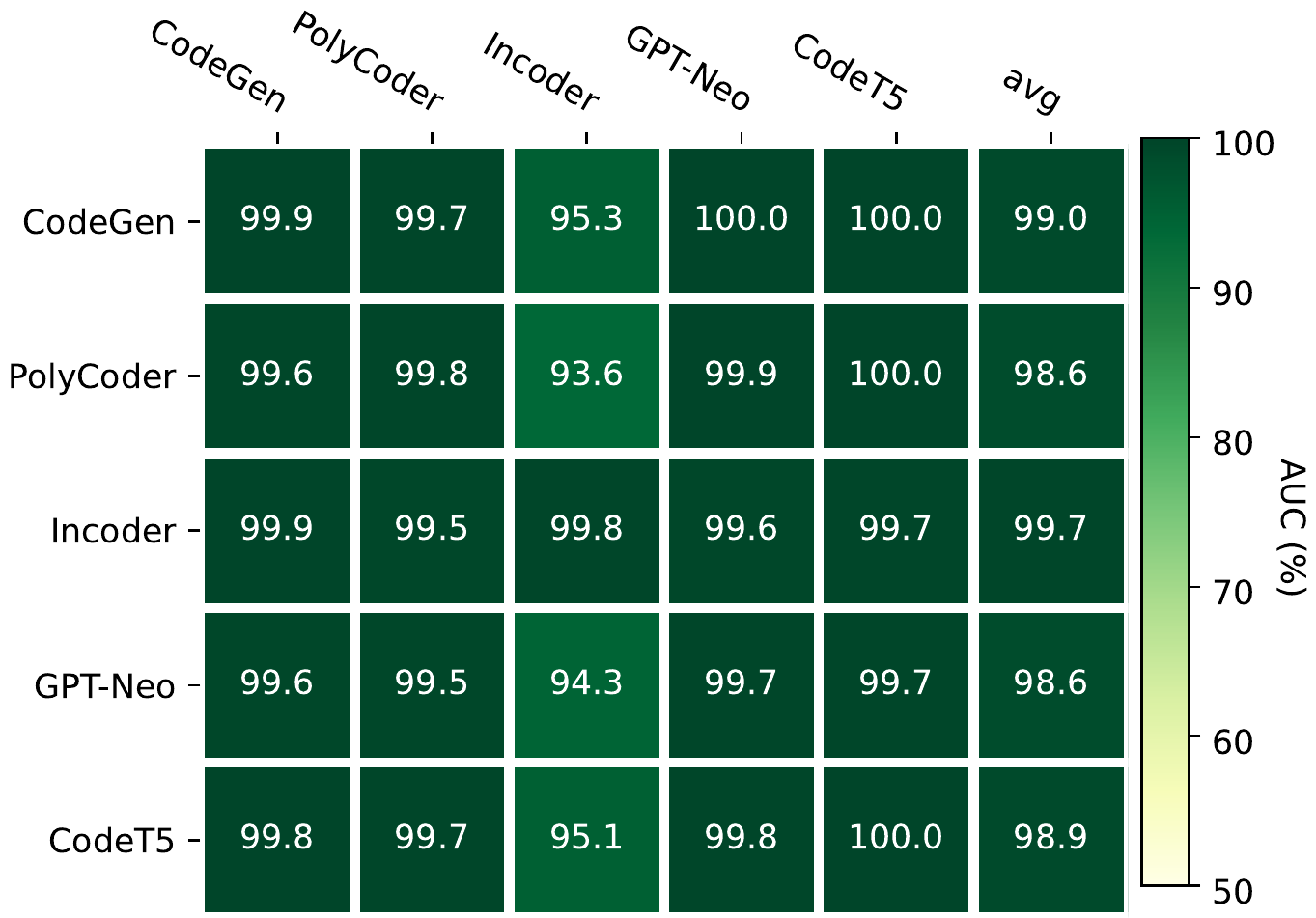}
\label{fig: detection_results_q1_apps}
\end{minipage}}
\vspace{-3mm}
\caption{\label{fig: detection_results_q1} The AUC scores of neural code detection on three datasets, (a) HumanEval, (b) MBPP, and (c) APPS. The PLG models on the y-axis are used to generate neural code for the training set, while models on the x-axis are used to generate neural code for evaluation (same for the figures in the following sections).
} 
\vspace{-3mm}
\end{figure*}

\subsection{Methodology}
\label{sec: neural code detection Methodology}

To achieve the two aforementioned goals, we propose to learn a detector to capture the common fingerprints among PLG models. 
The detector is intrinsically a function mapping any code snippet to a label, \textit{i.e.}, neural or human-written. Therefore, the training data should consist of neural and human-written code snippets. The human-written code can be directly obtained from the benchmark dataset (\textit{e.g.}, MBPP) while the neural code is generated by PLG models. More concretely, the learning pipeline contains the following four stages:

\noindent \textbf{Stage 1: Generate Neural Code.} 
The neural code is generated by inputting prompts to PLG models. 
Here we uniformly notate code snippets as $\mathcal{X}_k:=\{x_i^k\}_{i=1}^{N}$, where $k=0$ means human-written code and $k>0$ means neural code. 
To evaluate the generalization of the learned detector, we here only use one PLG model, \textit{i.e.}, $\mathbf{G}_k$ from a set of PLG candidates $\mathcal{G}:= \{\mathbf{G}_k\}_{k=1}^{K}$ to generate neural code in the training set and observe its detection effectiveness.

\noindent \textbf{Stage 2: Generate Labels.}
The second stage for training set preparation is to label the snippets of code. 
For each code snippet $x^k_i$, its label $y_i$ will be 0 if $k=0$ while 1 if $k > 0$. Given paired $x^k_i$ and $y_i$, our training set is denoted as $\mathcal{D}:= \{(x^k_i, y_{i})|i=1,\dots,N, k=0 \text{~or~}k\in \{1,2,...,K\} \}$.

\noindent \textbf{Stage 3: Learn the Detector.} 
With the training set $\mathcal{D}$ at hand, we can train a binary classifier $\mathcal{F}$ mapping a code snippet to a probabilistic vector $\mathcal{F}(x_i)$ over all possible classes. Training $\mathcal{F}$ is to solve the following optimization problem: $min_{\mathcal{F}}\Sigma_{x_i \in \mathcal{X}}L(\mathcal{F}(x_i), y_i)$, where $L(\mathcal{F}(x_i), y_i)$ is a loss function that measures the differences between the model output $\mathcal{F}(x_i)$ and the ground truth label $y_i$.

\noindent \textbf{Stage 4: Inference.} After the training stage, given a code snippet~$x$, the label of $x$ is $\operatorname{argmax}\mathcal{F}(x)$. In other words, we consider the value with the highest probability as the most possible class $\hat{y}$ for the source-unknown code snippet $x$. We note that our testing neural code snippets are generated not only by model $\mathbf{G}_k$ but also by models in $\mathcal{G}/\mathbf{G}_k$.

\subsection{Experimental Setup}
\label{sec: neural code detection Experimental Setup}

\noindent \textbf{Dataset Configurations.} In this experiment, we use three datasets, including the HumanEval, MBPP, and APPS datasets. We do not use the MXBP dataset because it does not contain human-written code solutions. Unless stated otherwise, we only use the test set of the APPS dataset, which contains 5,000 prompts and corresponding solutions. Our experimental results show that a size of around 500 prompts is sufficient for training a well-performed detector.
We randomly select 80\% prompts from each dataset to generate neural code. The generated code is combined with its corresponding human-written code snippets for training, while the remaining 20\% prompts are used in the inference stage for evaluation.

\noindent \textbf{Code Generation Models and Configurations.}
We choose the larger-size version of each code generation model by default, \textit{i.e.}, CodeGen-6B, InCoder-6B, PolyCoder-2.7B, GPT-Neo-2.7B, and CodeT5-base (220M). We choose CodeGen-6B instead of CodeGen-16B for sake of generation efficiency. For the sampling parameters for all PLG models, 
we use the $\text{softmax}(x/T)$ sampling with a small temperature parameter $T$ = 0.2, since a lower temperature (more greedy decoding) achieves better generation quality when only generating one code snippet~\cite{nijkamp2022codegen, fried2022incoder, xu2022systematic, chen2021evaluating, austin2021program, le2022coderl}.
Additionally, we use the most commonly used top-$p$ sampling (so called nucleus sampling)~\cite{holtzman2020curious} with $p$ = 0.95~\cite{nijkamp2022codegen, fried2022incoder, xu2022systematic, chen2021evaluating, wang2021codet5}.

\noindent \textbf{Detector Construction.} 
When building our detector, we fine-tune UniXcoder~\cite{guo2022unixcoder}, a state-of-the-art pre-trained programming language model to improve the effectiveness of neural code detection. UniXcoder is pre-trained on the CodeSearchNet dataset and an additional code dataset containing C, C++, and C\# programming language. Thus, UniXcoder can support nine languages: Java, Ruby, Python, Php, Javascript, Go, C, C++, and C\#. We use AdamW optimizer \cite{loshchilov2018decoupled} with learning rate $r=10^{-6}$.

\subsection{Results \& Analysis}
\label{sec: neural code detection Results and Analysis}
This section presents the main results of the detection performance. We use Area Under ROC Curve (AUC) as the evaluation metric since neural code detection is a binary classification task.

Figure \ref{fig: detection_results_q1} shows the performance of our neural code detector in \textit{in-domain} and \textit{cross-domain} settings. For the in-domain setting where training and testing neural code snippets are generated by the same model (\textit{i.e.}, results on diagonal entries), our detector achieves almost 100\% AUC scores on all three datasets. This indicates the existence of common fingerprints shared across PLG models, and such fingerprints are inherited by their generated code. Moreover, the results show that our detector can leverage these fingerprints to detect neural code generated by known models, achieving the first goal.

For the cross-domain setting where training and testing code snippets are generated by different generation models (\textit{i.e.}, results on off-diagonal entries), our detector achieves average AUC scores of $>$83\%, $>$93\%, and $>$98\% on three datasets, respectively. 
Furthermore, we evaluate the performance of our detector in the case that training and testing prompts are from different datasets. The results show that our detector achieves $>$80\% average AUC scores when trained with MBPP and tested on HumanEval and APPS (Detailed results are shown in Figure B1 in Appendix B). 
These results indicate that our neural code detector has good generalization ability, achieving the second goal of detecting neural code generated by unseen models.

We can also observe that our detector yields better performance on APPS. The reason is that APPS has a larger size of the training data than the others. Specifically, APPS has 4,000 neural code snippets and another 4,000 corresponding human-written code snippets for training. In contrast, MBPP only has $771 \times 2 = 1542$ snippets and HumanEval has even fewer ($131 \times 2 = 262$). 
Additionally, our additional experiments find that the performance of the detector will not drop significantly on MBPP and APPS when we manually decrease the size to at least $500 \times 2 = 1000$. Therefore, we recommend building the detector with as least 1,000 snippets to achieve better detection performance.

\noindent \textbf{Can Neural Text detector Handle Neural Code?}
The previous work~\cite{ippolito2020automatic} showed that their fine-tuned BERT-based classifier could outperform other methods in detecting neural texts. We adopted this well-performed neural text detector to detect neural code snippets generated by each PLG model. However, it yields AUC scores ranging from 53.3\% to 69.2\%, indicating the inability of the neural text detector to handle neural code.

Furthermore, we evaluate the detection performance when fine-tuning BERT~\cite{Devlin2019bert} using the training data $\mathcal{D}$ constructed from the MBPP dataset (Detailed results are shown in Figure B2 in Appendix B). We can find that this BERT-base detector achieves almost the same performance as our UniXcoder-based detector (Figure~\ref{fig: detection_results_q1_mbpp}) in the in-domain setting.
However, it shows significant performance decreases in generalization ability. For example, the AUC scores of the BERT-base detector are all greater than 95\% for the in-domain setting, while it drops to nearly 50\% when the testing code snippets are generated by GPT-Neo but the training data is generated by PolyCoder, InCoder, or CodeT5. This indicates that a purely natural language-based pre-trained model, such as BERT, lacks the ability to understand the distinct semantic and syntactic information of programming languages.
These results highlight the importance to explore the properties of PLG models for neural code attribution.

\begin{tcolorbox}
\begin{itemize} [leftmargin=*]
\item Extensive evaluations demonstrate the existence
of common fingerprints shared across different PLG models.
\item The inefficiency of the neural text detector highlights the critical need to investigate the features of PLG models and their generated neural code.
\end{itemize}
\end{tcolorbox}

\section{Challenges of Neural Code Attribution}
Section~\ref{sec: Neural Code Detection} shows that we can effectively distinguish neural code from human-written code by learning a binary classifier. However, this can only tell us the given neural code snippet is generated by \textit{some} models. In this section, we delve deeper to investigate the problem of neural code attribution (\textbf{RQ3}), \textit{i.e.}, attributing a given neural code snippet to \textit{the} model (\textit{e.g.}, a suspected model). 
Unfortunately, our empirical analysis and theoretical understanding demonstrate that it is NOT feasible to address this one-to-one attribution problem due to inherent challenges in neural code attribution.

\subsection{Empirical Analysis}
\label{sec: Empirical Analysis}

In this section, we propose two intuitive methods that attempt to solve the problem of neural code attribution. Our experimental results demonstrate the ineffectiveness of these two methods.

\subsubsection{One-Class Classification}
\label{One-Class Classification}
One may intuitively formulate the one-to-one attribution problem (where only one model is known) as a one-class classification problem (where only one class is known). Therefore, we can build a one-class classifier with neural code generated by the known model to detect neural code generated by unknown models. To examine whether the one-class classifier can effectively address this problem, we employ the widely used one-class support vector machine (SVM)~\cite{scholkopf1999support} with a Gaussian kernel.

The preparation of the training set is similar to Stage 1 described in Section~\ref{sec: neural code detection Methodology}. The difference is that we do not need to use human-written code here. As we cannot fine-tune UniXcoder in the one-class classification problem, we use it as a feature extractor to transform the raw code data into numerical features that can be processed by the one-class SVM. In this way, our one-class SVM model can also benefit from its code understanding ability. Following the common practice~\cite{Devlin2019bert, guo2022unixcoder}, we take the last hidden state at the final layer of UniXcoder as the embedding feature of code $x$.

\begin{figure}[tp]
\centering
\includegraphics[width=0.99\linewidth]{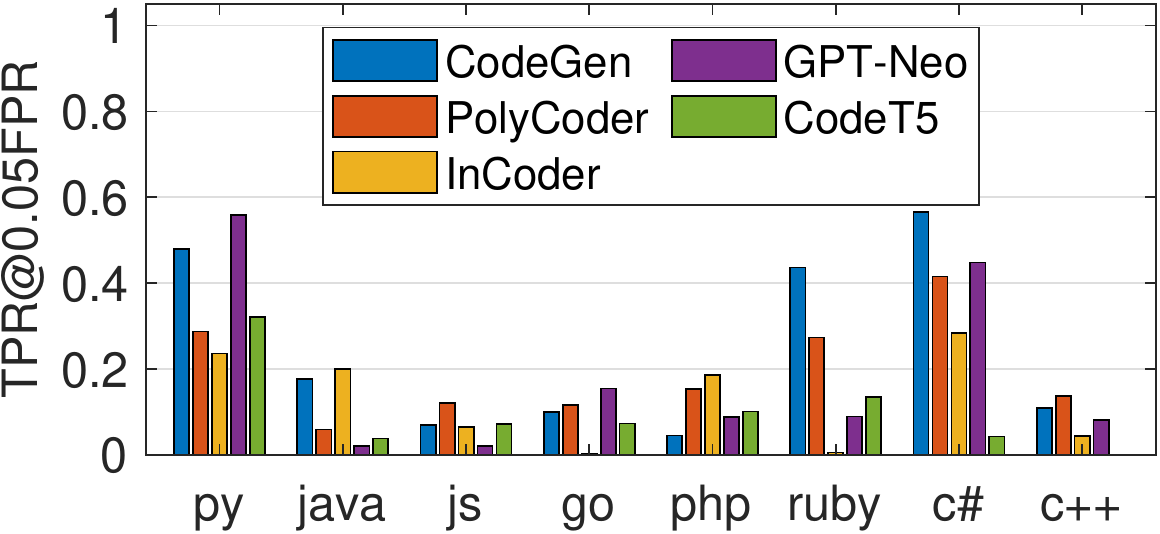}
\vspace{-1mm}
\caption{\label{fig: detection_results_q3_oneclass} Performance of attribution model using one-class SVM on different programming languages.}
\vspace{-3mm}
\end{figure}

We evaluate the performance of one-class SVM detectors for each PLG model on all programming language datasets except for C++ language because C++ is not supported by CodeT5. The results are reported in Figure~\ref{fig: detection_results_q3_oneclass}. It can be observed that the performance of using one-class SVM for neural code attribution is very poor in most cases. For example, the one-class SVM detector trained with code snippets generated by CodeGen yields true positive rates (TPRs) of only 6.9\%, 10\%, and 4.5\% on Java, Go, and PHP, respectively with a fixed false positive rate (FPR) at 5\%. Similar to the results for CodeGen, the average TPR @ 5\% FPR on all datasets for Polycoder,	InCoder, GPT-Neo, and CodeT5 are only 19.5\%, 12.7\%, 18.2\%, and 11.14\%, respectively. 
Figure~\ref{fig: mbpp_unixcoder-base-nine_mean_mbpp_t-SNE} illustrates the representation distributions of Python code snippets generated by different PLG models. We can observe that the output distributions of different PLG models are heavily overlapped, impeding the one-class SVM model from learning a clear support boundary.

\begin{figure}[tp]
\centering
\includegraphics[width=0.99\linewidth]{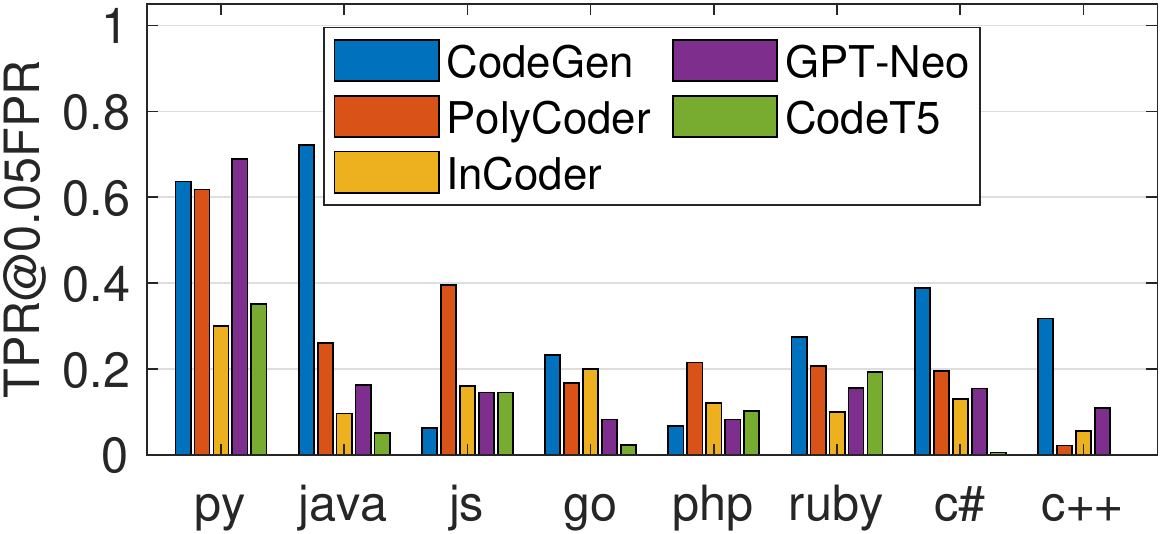}
\vspace{-1mm}
\caption{\label{fig: detection_results_q3_likelihood} Performance of attribution model using likelihood-based attribution on different programming languages.}
\vspace{-2mm}
\end{figure}

\subsubsection{Likelihood-based Attribution}
\label{sec: Likelihood-based Attribution}
The one-to-one attribution problem may also be addressed by comparing the likelihood that a code snippet is generated by a given PLG model, as the PLG model would give a higher likelihood to its own generated code snippets compared to the code snippets generated by other models. Similar to the likelihood ratio test in Section~\ref{sec: Training Usage Auditing Methodology}, we use the perplexity value of each sample to perform our likelihood-based attribution.

We report the results of the likelihood-based attribution on different programming language datasets in Figure~\ref{fig: detection_results_q3_likelihood}. We can observe that likelihood-based attribution cannot effectively determine whether a code snippet is generated by the given model. For example, although likelihood-based attribution achieves 72.1\% TPR @ 5\% FPR when distinguishing Java code snippets generated by CodeGen from those generated by the other models, it yields only 6.3\%, 23.2\%, and 6.7\% on Javascript, Go, and PHP, respectively. 
In most cases, the TPRs @ 5\% FPR of likelihood-based attribution is around 20\%. The average TPR @ 5\% FPR on all datasets for Polycoder, InCoder, GPT-Neo, and CodeT5 is only 25.9\%, 14.5\%, 19.7\%, and 12.4\%, respectively.

\begin{figure}[tp]
\centering
\includegraphics[width=0.65\linewidth]{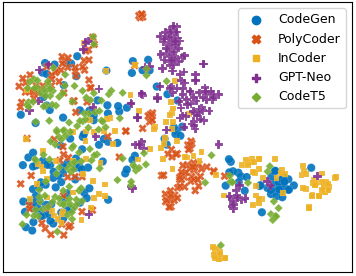}
\vspace{-1mm}
\caption{The representation distributions of Python code  generated by different PLG models. The representations are extracted by UniXcoder and projected by t-SNE\cite{van2008visualizing}.}
\label{fig: mbpp_unixcoder-base-nine_mean_mbpp_t-SNE}
\vspace{-1mm}
\end{figure}

\subsection{Understanding the Failure}
\label{sec: Understanding the Failure}
The empirical analysis in Section~\ref{sec: Empirical Analysis} has demonstrated the failure of using two intuitive methods to solve the problem of neural code attribution. In this section, we provide a theoretical analysis to prove that it is an impossible task to reliably attribute a single code snippet to a given PLG model. %, which essentially explains the failures of ineffective attempts in the previous section.
We formalize our \textbf{RQ3} as a single-instance goodness-of-fit hypothesis test~\cite{serra2019input, wang2020further}. Subsequently, we prove that \textbf{RQ3} as a single-instance distributional test is impossible, \textit{i.e.}, the test does no better than random guessing. This is done by generalizing theoretical results from previous works in OOD detection \cite{zhang2021understanding} to our neural code attribution setting.

For a given code snippet $x$ and a PLG model $\mathbf{G}$, the goal of \textbf{RQ3} is to determine whether $x$ comes from the model $\mathbf{G}$'s output distribution $\P$ or not. 
This can be formally defined as follows:
\begin{definition}
\label{definition: Single-instance Goodness-of-Fit Test}
(Single-Instance Goodness-of-Fit Test)
Formally, given an instance $x$, the test decides whether to reject the null hypothesis ($H_{0}$) that $x$ was drawn from the data distribution $\P$, in favor of an alternative hypothesis ($H_{1}$) that $x$ came from a distribution other than $\P$:
\begin{align}
    &H_{0}: x \sim \P,\\
    &H_{1}: x \sim \Q \in \mathcal{Q}, \P \notin \mathcal{Q}, 
\end{align}
where $\mathcal{Q}$ is a set of alternative distributions, \textit{i.e.}, the unknown output distributions of other PLG models except for the given model $\mathbf{G}$.
\end{definition}

Let $\phi(x)$ be a predetermined test statistic, which is a function of a single instance $x$. Based on the value of $\phi(x)$, one can decide whether to reject the null hypothesis or not.
The quality of a test is measured by the test power (\textit{i.e.}, TPR) under a pre-defined significance level ($\alpha$), also known as the Type I error or false positive rate.

The formalized single-instance goodness-of-fit test is a challenging statistical inference task given that there is no knowledge about the output distributions of other PLG models. As shown in Proposition~1 in the previous work~\cite{zhang2021understanding}, the single-instance distributional test formalized in Definition~\ref{definition: Single-instance Goodness-of-Fit Test} has test power equal to the false positive rate, which is equivalent to random guessing. 
Proposition~1 in \cite{zhang2021understanding} states that no test statistic can do well against all alternative data distributions. In other words, any test statistic $\phi(x)$ for single-instance distributional testing must trade off their test power against different unknown alternative distributions $\Q \in \mathcal{Q}$.

For further explanation, we use an intuitive example to elaborate on the impossibility of the single-instance distributional test. As aforementioned, we consider test statistics that directly utilize knowledge of the output distribution $\P$, which can be denoted as $\phi_{\P}: \mathcal{X} \rightarrow \mathbb{R}$. Suppose $\mathcal{X} \in \mathbb{R}^{d}$, then the test statistic $\phi_{\P}$ is a mapping function from $\mathbb{R}^{d} \rightarrow \mathbb{R}$. 
Therefore, the statistic value $\phi_{\P}(x)$ is a projection of $d$-dimensional distribution onto a one-dimensional space of the test statistic. 
Since a single marginal distribution cannot capture all differences between two multivariate distributions, the projections of $\P$ and $\Q$ on the test statistic cannot assess all their differences.

The above theoretical analysis suggests that, without additional knowledge about the output distributions of other PLG models, no method for one-to-one attribution can guarantee performance beyond random chance.
This indicates one-to-one attribution is an impossible task. Therefore, we propose two feasible alternative questions by relaxing the constraints on the number of candidate models and the number of neural code snippets, respectively. Specifically, we present to address \textit{attribution classification} (\textbf{RQ3.1}) which aims to classify a given neural code snippet to its source model among a set of candidate models (Section~\ref{sec: Attribution Classification}), and \textit{attribution verification}  (\textbf{RQ3.2}) which aims to verify whether a set of neural code snippets is generated by a given PLG model (Section~\ref{sec: Attribution Verification}).

\begin{tcolorbox}
\begin{itemize} [leftmargin=*]
\item Our empirical and theoretical analyses demonstrate that it is impossible to reliably attribute the generation of one code snippet to a specific model, shedding light on the challenges of dissecting neural code accountability.
\end{itemize}
\end{tcolorbox}

\section{Attribution Classification}
\label{sec: Attribution Classification}
In this section, we present our study of attribution classification, addressing the \textbf{RQ3.1}, \textit{i.e.}, given a neural code $x$ and $K$ candidate PLG models, can we single out the source model (among $K$ alternatives) that generated $x$? To answer this question, we learn a multi-class classifier to predict which PLG model generates the neural code snippet.
We begin with our intuition and goals. We then detail our method to build the multi-class classifier. Finally, we present the experimental setup and evaluation results.

\subsection{Intuition and Goals}
\label{sec: attribution classification Intuition and Goals}
We drive our intuition behind the attribution classification from the fact in NLG models that neural texts generated by NLG models carry subtle but discriminating features or fingerprints inherited from their source models and that these features can be utilized by machine learning (ML) algorithms for attribution~\cite{uchendu2020authorship, munir2021through, pan2020privacy}. Therefore, we believe that PLG models also hold the same properties, \textit{i.e.}, code snippets generated by PLG models inherit subtle but different fingerprints from their source PLG models. To this end, we perform an attribution classification approach aiming to achieve the following goals:

{$\vcenter{\hbox{\tiny$\bullet$}}$} 
\textbf{Identifying Source Models.} The primary goal of attribution classification is to predict which PLG model generates the neural code among a set of model candidates. This can verify whether the generated code carries distinguishing fingerprints inherited from their source models.

{$\vcenter{\hbox{\tiny$\bullet$}}$} 
\textbf{Generalization to different prompts.} In real-world settings, neural code snippets can be generated by prompts from different datasets. As a result, the distribution of the prompts may also be different. Therefore, the attribution classifier should have good generalization performance when neural code snippets for training and testing are generated by prompts from different datasets.

\subsection{Methodology}
\label{sec: attribution_classification_methodology}
We propose a multi-class classification learning method to achieve the above-mentioned two goals for attribution classification.
Similar to neural code detection, the pipeline consists of four stages. 

\noindent \textbf{Stage 1: Generate Neural Code.}
We use a set of prompts to query each model in $\mathcal{G}:= \{\mathbf{G}_k\}_{k=1}^{K}$ to generate a set of neural code $\mathcal{X}_k:=\{x_i^k\}_{i=1}^N$.

\noindent \textbf{Stage 2: Generate Label.} 
For each $x_i^k$ in $\mathcal{X}_k$, we assign $y_i=k$ as its label. We construct our training set as $\mathcal{D}:= \{(x^k_i, y_{i})|i=1, \dots, N, k=1, \dots, K\}$ .

\noindent \textbf{Stage 3: Learn the Multi-class Classifier.}
We train a multi-class classifier $\mathcal{F}$ with the constructed dataset $\mathcal{D}$ for addressing the attribution classification problem.

\noindent \textbf{Stage 4: Inference.} After the training stage, we can evaluate the learned multi-class classifier to predict the most possible source model of a given testing code snippet $x$, which may be generated by any model $\mathbf{G}_k$ in $\mathcal{G}$.

\subsection{Experimental Setup}
\label{sec: attribution classification Experimental Setup}
Since our attribution classification task is very similar to neural code detection in Section \ref{sec: Neural Code Detection}, we mainly describe the different parts of the experimental setup for attribution classification. 

\noindent \textbf{Dataset Configurations.} 
We use MBPP and MBXP to evaluate the performance of our attribution classifier for multiple programming languages, \textit{i.e.}, MBPP for Python, MBXP for Java, Javascript, Go, PHP, Ruby, C\#, and C++. 
Similar to neural code detection, we randomly split the prompts with an 80-20 ratio to generate neural code snippets for training and testing, respectively.

\noindent \textbf{Code Generation Models.}
As shown in Table \ref{table: summary of code generation models}, different types of PLG models are trained on different sets of programming languages. Therefore, we do not include CodeT5 for MBXP-CPP dataset since the training dataset of CodeT5 does not contain any code in the C++ program language. 

\subsection{Results \& Analysis}
\label{sec: attribution_classification_Result_Analysis}

\begin{table}[]
\centering
\caption{\label{table: detection_results_q2} Performance comparison of attribution classification across the different programming languages.}
\vspace{-2mm}
\resizebox{0.98\linewidth}{!}{
\begin{threeparttable}
\begin{tabular}{cccccccccc}
\toprule
& \textbf{py} & \textbf{java} & \textbf{js} & \textbf{go} & \textbf{php} & \textbf{ruby} & \textbf{c\#} & \textbf{c++} & \textbf{all\textsuperscript{*}} \\
\midrule
UniXcoder-based & \textbf{0.994}        & \textbf{0.955}          & \textbf{0.900}         & \textbf{0.872}        & \textbf{0.941}        & \textbf{0.970}         & \textbf{0.893}         & \textbf{0.915}        & \textbf{0.932}        \\
\midrule
CodeBERT-based  & 0.990        & 0.889         & 0.873       & 0.751       & 0.905        & 0.966          & 0.789        & 0.850         & 0.891\\
\midrule
BERT-based  & 0.914        & 0.882         & 0.826       & 0.719       & 0.882        & 0.953          & 0.775        & 0.829         & 0.869\\
\midrule
DL-CAIS~\cite{abuhamad2018large}  & 0.711 & 0.844	& 0.720	& 0.690	& 0.769	& 0.863	& 0.742	& 0.786 & - \\
\midrule
PbNN~\cite{bogomolov2021authorship}  & 0.702 & 0.861 & - & - & - & - & - & 0.750 & - \\
\midrule
PbRF~\cite{bogomolov2021authorship}  & 0.812 & 0.906 & - & - & - & - & - & 0.877 & -\\
\bottomrule
\end{tabular}
\begin{tablenotes}
\item[\textsuperscript{*}] the attribution classifier is trained and evaluated in the multi-lingual setting.
\end{tablenotes}
\end{threeparttable}
}
\vspace{-2mm}
\end{table}

We now present the main results of the  attribution classification. Table~\ref{table: detection_results_q2} shows that our attribution classification obtains good performance, achieving the first goal. For instance, in the mono-lingual attribution classification setting where our attribution classifier is trained and tested on mono-lingual neural code snippets, our attribution classifier achieves accuracy scores from 87.2\% to 99.4\% in different programming languages. In the multi-lingual setting where our attribution classifier is trained and tested on neural code snippets in all languages, our attribution classifier still maintains a remarkable performance, achieving an accuracy score of 93.4\%. These results indicate that code snippets generated by PLG models inherit subtle but unique fingerprints from their source PLG models, leading to the success of our attribution classifier in identifying the fingerprints. 

Furthermore, we evaluate our attribution classifier's generalization ability in the case that training and testing prompts are from different datasets.
We find that our classifier achieves an accuracy score of 92.1\% when training with APPS and testing on MBPP. 
When we compared it to the classifier accuracy of both training and testing on MBPP, \textit{i.e.}, the 99.4\% of Python result in Table~\ref{table: detection_results_q2}, 92.1\% is still a comparably satisfied accuracy. This indicates that our attribution classifier has a good generalization ability to neural code snippets that are generated by prompts from different distributions, achieving the second goal.

\noindent \textbf{Comparison with Authorship Attribution of Human-written Code.} 
The task of our attribution classification for neural code is very similar to a well-studied area, source code authorship attribution~\cite{abuhamad2018large, bogomolov2021authorship}, which aims to identify the author of a given piece of code from a predefined set of authors. 
Here, we compare our attribution classification method with three human-written source code authorship attribution baselines, including DL-CAIS~\cite{abuhamad2018large}, PbNN~\cite{bogomolov2021authorship}, and PbRF~\cite{bogomolov2021authorship}. These methods are the most recent and state-of-the-art ones, and they are open-sourced. Since PbRF and PbNN require a specific parser for each target language, their implementation only supports Python, Java, C, and C++. 

As shown in Table~\ref{table: detection_results_q2}, we can observe that our proposed attribution classification method (UniXcoder-based one) outperforms all three human-written source code authorship attribution methods. For example, our proposed method achieves an accuracy score of 99.4\% in Python while DL-CAIS, PbNN, and PbRF yield only 77.1\%, 70.2, and 81.2\%, respectively. 
The premise for the success of authorship attribution is that each author has a unique coding style, such as lexical, syntactic, and semantic patterns~\cite{li2022ropgen}. Our experimental results show that these authorship attribution methods do not work effectively for neural code attribution. We conjecture that this is because PLG models are trained on large corpora from different authors thus there may be no clear-cut stylometric differences in neural code generated by PLG models.

\noindent \textbf{Impact of Pre-trained Code Understanding Model.}
The performance of our attribution classifier relies on the code understanding ability of the pre-trained programming language model.
Here, we explore the impact of the choice of code understanding models on attribution performance. The results are shown in Table~\ref{table: detection_results_q2}. 
Similar to our findings in Section~\ref{sec: neural code detection Results and Analysis}, we can observe that the attribution classifier built on a purely natural language-based pre-trained model, like BERT~\cite{Devlin2019bert}, yields the worst performance compared to the classifiers built on pre-trained models for programming languages.

Additionally, UniXcoder is preferable given both UniXcoder and CodeBERT are specifically pre-trained on source code. This is because, in addition to semantic information of code, UniXcoder can leverage syntax information of abstract syntax tree (AST) to enhance code representation learning.
Moreover, we should point out that the CodeBERT-based attribution classifier yields comparable performance in six programming languages, \textit{i.e.}, Python, Java, Javascript, Go, Php, and Ruby. However, there are large accuracy drops in the other two programming languages, \textit{i.e.}, C\# and C++. We suppose the reason is that CodeBERT is trained on the CodeSearchNet dataset that only contains the first six programming languages. As a result, it does not have a good understanding ability of other languages, such as C\# and C++, leading to a significant accuracy drop compared to our proposed UniXcoder-based classifier. 

\noindent \textbf{Distinguishing PLG Models in a Family.} 
In the above evaluation, our attribution classifier aims to learn and distinguish fingerprints from different PLG families, where they differ from each other in many ways.
For example, as for training objectives, CodeGen, PolyCoder, and GPT-Neo utilize a left-to-right (causal) autoregressive language modeling objective~\cite{brown2020language, chen2021evaluating}, InCoder adopts a causal masking objective~\cite{aghajanyan2022cm3}, and CodeT5 includes masked span prediction~\cite{raffel2020exploring} and denoising sequence reconstruction~\cite{lewis2020bart} objectives. In terms of model architectures, CodeGen, PolyCoder, GPT-Neo, and InCoder are decoder-only models while CodeT5 is an encoder-decoder model. 

Here, we explore whether our attribution classification can distinguish fingerprints of more similar models in the same family, \textit{i.e.}, models that are different in model size but designed by the same algorithm. In the experiments, we use models in the CodeGen family as it provides models trained in four sizes, 350M, 2.7B, 6B, and 16B. Particularly, we conduct this experiment on the MBPP dataset with CodeGen-Mono-350M, CodeGen-Mono-2.7B, CodeGen-Mono-6B, and CodeGen-Mono-16B. The result shows that our attribution classifier achieves a good accuracy score of 83.87\%, much higher than~the four-class random guess accuracy of 25\%. This result demonstrates that our attribution classification can effectively learn and distinguish fingerprints of similar PLG models that are designed by the same algorithm.

\subsection{Discussion}
\label{sec: attribution_classification_discussion}
In the previous evaluations, we demonstrate the subtle but distinctive fingerprints in PLG models, which are effectively captured by our attribution classifier. Here, we aim to visualize the fingerprints of different PLG models. 

Similar to source code authorship attribution~\cite{li2022ropgen, abuhamad2018large, bogomolov2021authorship}, we can treat each of our PLG models as a code author. We suppose that neural code snippets generated by different PLG models may have different coding style patterns which reflect the unique fingerprints of PLG models. To verify it, we exploit the model interpretability tool, Captum~\cite{kokhlikyan2020captum} which provides state-of-the-art methods (\textit{e.g.}, Integrated Gradients~\cite{sundararajan2017axiomatic}) to explore which coding style patterns are contributing to the output of the attribution classifier.

We query each model to generate a code snippet from the same prompt that is randomly selected from MBPP. We then use Captum to compute the importance score (ranging from $-1$ to $1$) of each token. If a token has a large importance score, this token positively contributes to the classification.
For example, as shown in Figure B3 in Appendix B, the token ``print'' is denoted negatively contributing to the classification. This is a reasonable result because ``print'' is broadly used in most PLG models and cannot be used as a feature to distinguish PLG models. In contrast, the ``Parameters'' and ``Arguments'' are considered as positive tokens. This is because although they all mean defining the parameters, PolyCoder uses ``Parameters'' while InCoder uses ``Arguments''. Therefore, we can use this feature to help distinguish PolyCoder and InCoder. In summary, our case study finds that when distinguishing neural code snippets generated by different PLG models, the classification model relies on some coding style patterns such as function names, variable names, and local comment format in a function.

\begin{tcolorbox}
\begin{itemize} [leftmargin=*]
\item By relaxing the constraints on the number of candidate models, we propose attribution classification to predict the source model of a neural code with high accuracy. 
\item Extensive evaluations demonstrate different PLG models have distinctive fingerprints which are inherited by their corresponding generated neural code.
\item We analyze the distinct fingerprints of different PLG models from the perspective of coding style using the model interpretability tool.
\end{itemize}
\end{tcolorbox}

\section{Attribution Verification}
\label{sec: Attribution Verification}

In this section, we present our study of attribution verification, addressing the \textbf{RQ3.2}, \textit{i.e.}, can we verify whether a set of neural code snippets is generated by a given PLG model? To answer this question, we formulate attribution verification as a two-sample testing problem. We begin with our intuition and goals. We then detail our method based on two-sample hypothesis testing. Finally, we present the experimental setup and evaluation results.

\subsection{Intuition and Goals}

Our intuition behind the attribution verification is that the output distributions of different PLG models are different, otherwise, we cannot learn an effective attribution classifier. However, these output distributions are heavily overlapped, leading to the failure of the single-instance attribution as is discussed in Section~\ref{sec: Understanding the Failure}. Therefore, we here relax the constraint of one testing code snippet and propose to perform a hypothesis test to verify whether the distribution of the testing code snippets is the same as that of neural code snippets generated by the given model. If yes, the neural code snippets are generated by the given model. We aim to achieve the following goals in attribution verification:

{$\vcenter{\hbox{\tiny$\bullet$}}$} 
\textbf{Verifying Attribution.} The primary goal is to verify whether a set of neural code snippets is generated by the given PLG model. 

{$\vcenter{\hbox{\tiny$\bullet$}}$} 
\textbf{Generalization to different datasets.} The attribution verification method should have good generalization ability to neural code snippets that are generated by prompts from different distributions.

\subsection{Methodology}
Suppose we have a PLG model $\mathbf{G}$ and observe a collection of code snippets $\mathcal{X}^{\prime}:=\{x_{i}^{\prime}\}_{i=1}^{n}$ that are i.i.d. instances from some distribution $\{x_{i}^{\prime}\}_{i=1}^{n} \sim p(x^{\prime}|\mathbf{G}^{\prime})$, where $\mathbf{G}^{\prime}$ is an unknown PLG model. The goal is to verify whether $\mathcal{X}^{\prime}$ was drawn from $p(x|\mathbf{G})$.

To achieve that goal, we use two-sample testing techniques \cite{gretton2012kernel, lloyd2015statistical, liu2020learning}. More specifically, we have a set of testing instances (\textit{i.e.}, neural code snippets) $\{x_{i}^{\prime}\}_{i=1}^{n}$ to be verified, and we can generate a set of neural code snippets from the model $\mathbf{G}$, $\{x_{i}\}_{i=1}^{m} \sim p(x|\mathbf{G})$. Then we can test whether these two sets of code data are drawn from the same distribution (\textit{i.e.}, generated by the same model) or not.

Now we can formulate our hypothesis test as a two-sample testing problem. Suppose we have instances $\mathcal{X}:=\{x_i\}_{i=1}^{m}$ and $\mathcal{X}^{\prime}:=\{x_{i}^{\prime}\}_{i=1}^{n}$  drawn i.i.d. from distributions $\P$ and $\Q$, respectively. The two-sample testing assesses whether distributions $\P$ and $\Q$ are different based on the instances drawn from each of them.

A well-known test statistic for addressing this problem is the Maximum Mean Discrepancy ($\MMD$) \cite{gretton2012kernel}:
\begin{align}
    \MMD^{2}(\P,\Q) &= \sup_{f \in \mathcal{H}} \left| \mathbb{E}_{x\sim \P}[f(x)] - \mathbb{E}_{x^{\prime}\sim \Q}[f(y)] \right| \\
    &= \Vert\mu_{\P} - \mu_{\Q}\Vert_{\mathcal{H}}^{2}, 
\end{align}
where $f$ is a function of the reproducing kernel Hilbert space (RKHS) $\mathcal{H}$ associated with a positive definite kernel $k$. In practice, a popular option is to use the Gaussian kernel $k(x, x^{\prime}) = \exp(-\Vert x - x^{\prime} \Vert ^{2}_{2}/\gamma^{2})$ with bandwidth parameter $\gamma$. Then $\MMD$ is the distance between the mean embedding of the distribution $\P$ and $\Q$ in RKHS. We denote the unbiased estimation of the $\MMD$ statistic as $\widehat{\MMD}_u^2(\P,\Q)$, which can be computed by
{\footnotesize \begin{align}
\frac{1}{m^2}\sum^m_{i,j=1;i\neq j}k(x_i,x_j) - \frac{2}{mn}\sum^m_{i,j=1}k(x_i,x_j^{\prime}) + \frac{1}{n^2}\sum^m_{i,j=1;i\neq j}k(x_i^{\prime},x_j^{\prime}).\nonumber
\end{align} }

\noindent \textbf{Estimating the Null Distribution.}
It is complicated to build a test by directly estimating the distribution of $\MMD$ statistic in the null hypothesis \cite{liu2020learning}. We instead use the permutation test \cite{sutherland2017generative} to estimate the null distribution. This is a simpler and faster procedure if implemented carefully. According to previous work~\cite{fernandez2008test}, under the null hypothesis, the instances in $\mathcal{X}$ and $\mathcal{X}^{\prime}$ are exchangeable. Therefore, we can estimate the distribution of the $\MMD$ statistic under the null hypothesis by repeatedly calculating it with the instances randomly assigned to $\mathcal{X}$ and $\mathcal{X}^{\prime}$ each time.

\noindent \textbf{Asymptotic and Test Power of MMD.} 
In this part, we discuss the relationship between sample size and test power. Generally, a larger sample size results in higher test power of $\MMD$ test with a fixed $\alpha$. In other words, the TPR of our attribution verification will approach 100\% with a fixed FPR if we have sufficient code snippets. 
This is theoretically guaranteed by the asymptotics of $\MMD$ as presented below:
\begin{prop}[Asymptotics of $\widehat{\MMD}_u^2$ under $H_1$] \label{prop:asymptotics}
Under the alternative, $H_1 : \P \ne \Q$,
a standard central limit theorem holds
\cite[Section 5.5.1]{serfling}:
\begin{gather*}
\sqrt{n}(\widehat{\MMD}_u^2 - \MMD^2) \overset{\textbf{d}}{\to} \N(0, \sigma^2)\\
\sigma^2 := 4 \left( \E[h_{12} h_{13}] - \E[h_{12}]^2 \right), 
\end{gather*}
where $h_{12}$, $h_{13}$ refer to $h_{ij}$ below:
\begin{align*}
    h_{ij} :=
    k(X_i, X_j)
    + k(X^{\prime}_i, X^{\prime}_j)
    - k(X_i, X^{\prime}_j)
    - k(X^{\prime}_i, X_j).
\end{align*}
\end{prop}

Proposition~\ref{prop:asymptotics} proves that $\sqrt{n}(\widehat{\MMD}_u^2 - \MMD^2)$ converges in distribution to a normal distribution $\N(0, \sigma^2)$. Thus, the test power of $\MMD$ converges to a percentile of the standard normal distribution:
\[
    {\Pr}_{H_1}\left( n \widehat{\MMD}_u^2 > r \right)
    \overset{\textbf{d}}{\to} \Phi\left( \frac{\sqrt{n} \MMD^2}{\sigma} - \frac{r}{\sqrt{n} \, \sigma} \right)
.\]
The approximate test power can be confirmed by using the rejection threshold $r$ that can be determined by the permutation test mentioned above.
According to \cite{sutherland2017generative,liu2020learning} and  Proposition~\ref{prop:asymptotics}, 
it is also clear that the $r$ will converge to a constant, and $\MMD$, $\sigma$ are also constants.
Therefore, for a reasonably large $n$, the test power is dominated by the first term, 
\textit{i.e.}, $\sqrt{n}\MMD^2/\sigma$. Namely, for a reasonably large $n$, test power increases when $n$ increases.

\subsection{Experimental Setup}

Like our previous experiments in Section~\ref{One-Class Classification}, we utilize the pre-trained UniXcoder model to pre-process code data and obtain their embeddings as numerical features. 
We evaluate the test power (TPR) on 100 random subsets which are selected from the neural code datasets used in Section~\ref{sec: Attribution Classification}. We repeat this full process 10 times and report the average test power (TPR) of each test. When implementing the permutation test, we set the repeat times as 200 to estimate the distribution of $\MMD$ under the null hypothesis.

\begin{figure*}[tp]
\centering
\includegraphics[width=0.95\linewidth]{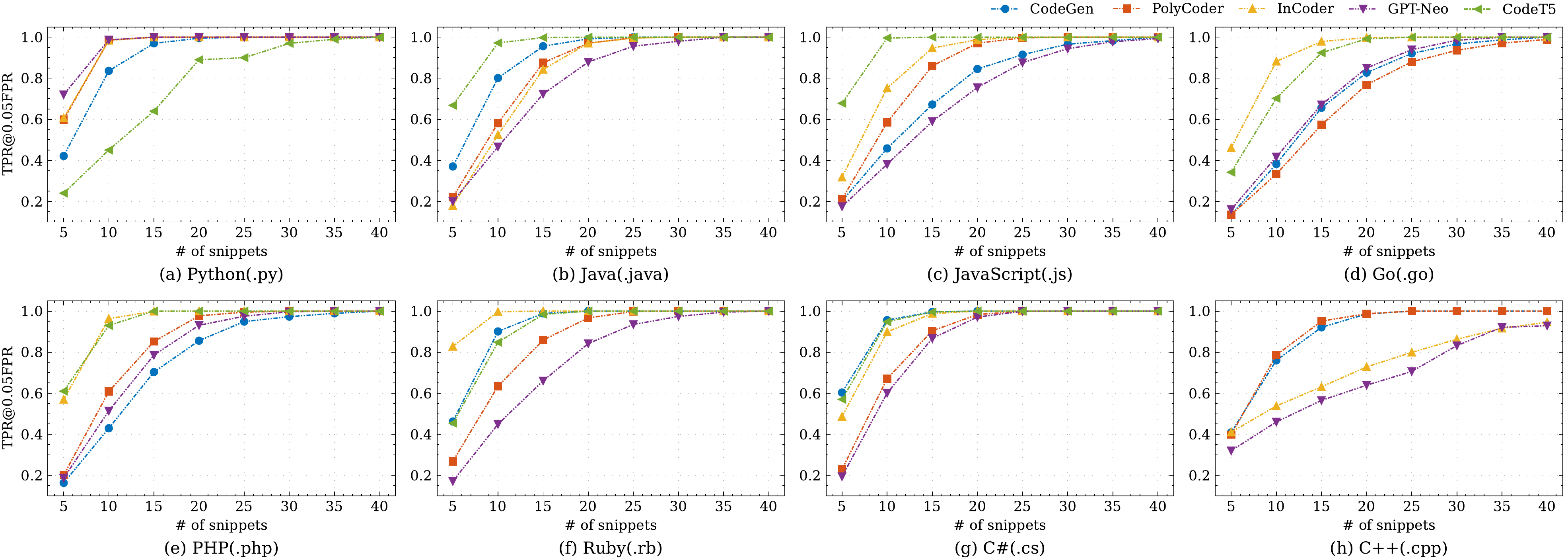}
\vspace{-1mm}
\caption{\label{fig: detection_results_q3_mmd} Performance of attribution verification on different programming languages under different numbers of available testing code snippets.}
\vspace{-1mm}
\end{figure*}

\begin{figure}[htp]
\centering
\subfigure[]{
\label{fig: detection_results_q3_mmd_ablation_apps} 
\begin{minipage}[t]{0.23\textwidth}
\centering
\includegraphics[width=0.99\textwidth]{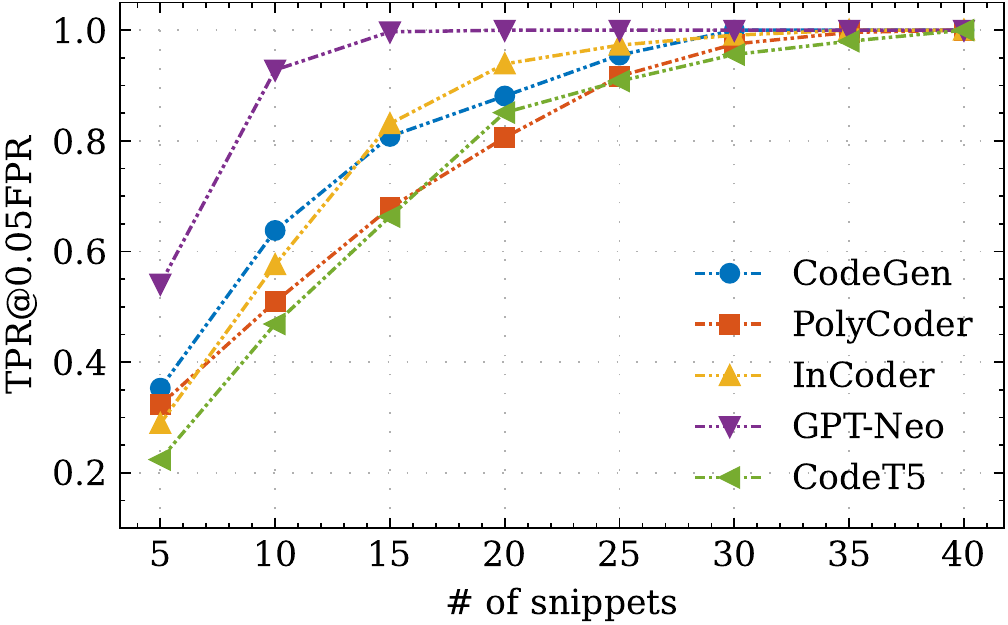}
\end{minipage}}
\subfigure[]{
\label{fig: detection_results_q3_mmd_ablation_codegen}
\begin{minipage}[t]{0.23\textwidth}
\centering
\includegraphics[width=0.99\textwidth]{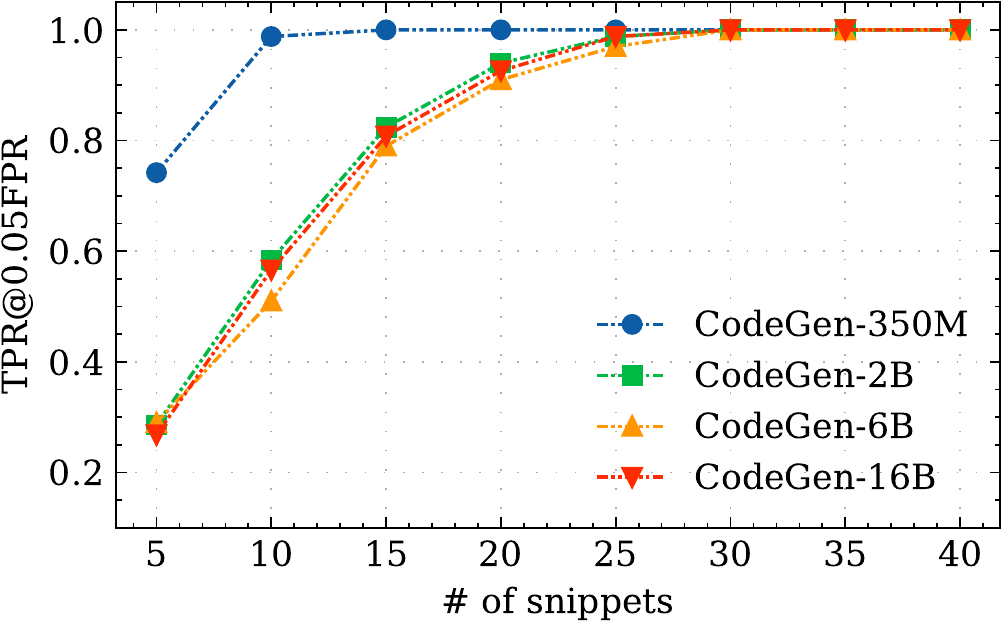}
\end{minipage}}
\vspace{-1mm}
\caption{(a) Performance of attribution verification when code snippets in $\mathcal{X}$ and $\mathcal{X}^{\prime}$ are generated with prompts from APPS and MBPP, respectively. (b) Performance of attribution verification when code snippets are generated by similar PLG models.}
\vspace{-1mm}
\end{figure}

\subsection{Results \& Analysis}

We now present the main results of the attribution verification. 
Figure~\ref{fig: detection_results_q3_mmd} illustrates the results of attribution verification using two-sample testing on different programming languages under different numbers of available testing instances. As expected, the number of available testing instances plays an important role in attribution verification, and the TPRs of attribution verification for all PLG models become larger with the increase in the number of testing code snippets. For example, we can observe that, in most cases, our attribution verification model achieves $>$80\% TPR @ 5\% FPR when the number of available testing instances increases to 20 or more, while the TPRs approach nearly 100\% when there are more than 30 testing instances. This result is consistent with the asymptotics of the $\MMD$ statistic. In other words, given sufficient testing instances, the TPR of our two-sample testing based attribution verification can approach 100\%.

\noindent \textbf{Generalization to Different Prompt Datasets.}
To evaluate the generalization ability of our proposed attribution verification method, we consider the scenario where the code snippets in $\mathcal{X}$ and $\mathcal{X}^{\prime}$ are generated from different prompt datasets. In the evaluation, we utilize the MBPP and APPS datasets to generate code snippets for $X$ and $\mathcal{X}^{\prime}$, respectively. We report the experimental results in Figure~\ref{fig: detection_results_q3_mmd_ablation_apps}. We can observe that our attribution verification achieves comparable performance when the number of testing code snippets is larger than 15, compared to the results in Figure~\ref{fig: detection_results_q3_mmd}(a). For example, the TPR @ 5\% FPR on PolyCoder drops 30\% when the number of testing code snippets is only 5, but it goes to 94.1\% when there are 15 testing code snippets, which is only a 5.9\% drop in TPR. Additionally, similar to the results on the in-distribution prompt dataset, our attribution verification method achieves $>$80\% TPR @ 5\% FPR on all five models when the testing code snippets are from a different prompt dataset as the number of testing code snippets increases to 20, while the TPRs on five models are all $>$95\% when the number of testing instances is 30 or more. These results demonstrate the  generalization ability of our proposed attribution verification method.

\noindent \textbf{Distinguishing Models in a Family.} 
Similar to the evaluation in Section~\ref{sec: attribution_classification_Result_Analysis}, we evaluate the performance of our attribution verification on distinguishing similar models that are different only in model size but designed by the same algorithm. We use four CodeGen models with different model sizes. Figure~\ref{fig: detection_results_q3_mmd_ablation_codegen} illustrates the verification results under different numbers of available testing instances. We can observe that our attribution verification achieves $>$90\% TPR @ 5\% FPR for all four models when the number of available testing instances increases to 20. Moreover, we can find that the TPR of CodeGen-mono-350M is much higher than those of the other three models. For example, the attribution verification of code snippets generated by CodeGen-mono-350M yields 74.2\% TPR @ 5\% FPR, even when there are only 5 testing code snippets. This means that the code snippets from CodeGen-mono-350M can be easily distinguished from those from the other models. We speculate the reason is that the poor code generation performance of CodeGen-mono-350M makes its output distribution lying relatively far away from those of other larger models.

\begin{tcolorbox}
\begin{itemize} [leftmargin=*]
\item By relaxing the constraints on the number of neural code snippets, we propose MMD-based attribution verification to reliably verify whether this set of neural code snippets is generated by a given PLG model.
\item We theoretically prove that our attribution verification can approach a 100\% TPR with a fixed FPR when we have sufficient neural code snippets.
\item Extensive evaluations on five PLG models and eight programming languages suggest that around 30 neural code snippets are sufficient for reliable verification.
\end{itemize}
\end{tcolorbox}

\section{Related Work}
\label{sec: Related Work}
In this section, we review the related work in terms of four aspects: language models of code, training data audit, detection and attribution of neural text, as well as authorship attribution of human-written source code. 

\subsection{Language Models of Code}
Inspired by the great success of large language models like GPT-2~\cite{radford2019language} for natural languages (NL), language models of programming language (PL) have recently shown tremendous promise in code understanding and generation tasks. 
The left-to-right nature of decoder-only models makes them effective for program synthesis tasks. Therefore, this line of research has seen huge progress with many recently proposed models such as CodeGPT~\cite{lu2021codexglue}, CodeX\cite{chen2021evaluating}, PolyCoder~\cite{xu2022systematic}, GPT-Neo~\cite{black2021gpt}, AlphaCode~\cite{li2022competition} and CodeGen~\cite{nijkamp2022codegen}.
Different from decoder-only models, encoder-only models (\textit{e.g.}, CodeBERT~\cite{feng2020codebert}, CuBERT~\cite{kanade2020learning}, and DISCO~\cite{ding2022towards}) commonly utilize a bidirectional training objective. This makes them suitable for producing an effective representation of the whole code snippet for downstream code-related understanding tasks. 
Encoder-only models and decoder-only models favor understanding and generation tasks, respectively. 
In contrast, encoder-decoder models can well support both code understanding and generation tasks. For example, PLBART~\cite{ahmad2021unified} is based on BART model~\cite{lewis2020bart} while \textsc{PyMT5}~\cite{clement2020pymt5} and CodeT5~\cite{wang2021codet5} explores the T5 framework~\cite{raffel2020exploring} for programming language.

\subsection{Training Data Audit} 

The aim of the training data audit is to determine if a data instance or set was used in training the model. 
A way is membership inference, which predicts whether or not a single data instance is used to train a target model.
Most of the existing studies on membership inference focus on classification models~\cite{shokri2017membership, ye2022enhanced, carlini2022membership, hu2022membership}. Only a few studies pay attention to language generation models~\cite{song2019auditing, hisamoto2020membership}. Song \textit{et al.}~\cite{song2019auditing} mount membership inference on LSTM-based text-generation models while Hisamoto \textit{et al.}~\cite{hisamoto2020membership} apply it to transformer-based sequence-to-sequence models. However, they either merely focus on shallow and simple models~\cite{song2019auditing} or yield poor performance on large transformer-based language models~\cite{hisamoto2020membership}. Moreover, they require to train multiple shadow models to mimic the behavior of the target model on member and non-member data instances. 
Recently, dataset watermarking has been proposed for data ownership verification in image~\cite{sablayrolles2020radioactive} and code~\cite{sun2022coprotector} domains. Sun \textit{et al.} propose an approach to trace the usage of open-source code in PLG models via dead-code insertion~\cite{sun2022coprotector}. However, the inserted dead code lacks stealthiness and could be easily identified by human inspection or filtered out by static code analysis tools.

\subsection{Detection and Attribution of Neural Text}
It has been played a lot of attention in developing ML-based approaches to distinguish between synthetic and human-written texts~\cite{gehrmann2019gltr, zellers2019defending, ippolito2020automatic}. He \textit{et al}. propose a comprehensive evaluation framework across different detection methods~\cite{he2023mgtbench}. 
Furthermore, given a neural text, recent work has further attempted to attribute the neural text to its source LM~\cite{uchendu2020authorship, pan2020privacy, munir2021through}.
In general, due to domain specificity, these existing methods that focus on natural languages are not directly applicable to programming languages, leaving the detection and attribution of neural code as an important but unexplored area. 
% \textbf{Language Watermarking.}
Watermarking methods for natural language is another technique to solve this attribution problem via marking and tracing text provenance. Recently, Abdelnabi and Fritz ~\cite{abdelnabi2021adversarial} propose an encoder-decoder network for text watermarking to replace the unobtrusive words in the original text with other inconspicuous words or symbols. However, such replacements still debilitate the logical and semantic coherence of the modified phrases~\cite{yang2022tracing}. Additionally, their dataset-specific model has poor transferability on text content with different text styles or data domains.

\subsection{Authorship Attribution of Human-written Source Code}
There is a rich body of literature on authorship attribution of human-written source code that aims to identify the author of a given piece of code from a predefined set of authors~\cite{abuhamad2018large, bogomolov2021authorship}. We discuss a few classic papers here. 
Caliskan \textit{et al.}~\cite{caliskan2015anonymizing} built a code stylometry feature set to represent coding style and learned a random forest for authorship attribution. Abuhamad \textit{et al.}~\cite{abuhamad2018large} trained fully-connected layers with coding style features extracted by TF-IDF. Bogomolov~\textit{et al.}~\cite{bogomolov2021authorship} utilized a pre-trained code representation model~\cite{alon2019code2vec} to extract deep features and predict the author of the snippet. 
Our study fills the gap in exploring the properties of neural code generated by PLG models and providing comprehensive analysis to trace their accountability, which differs from the aforementioned studies in authorship attribution.

\section{Conclusion}
This paper presents the first comprehensive study on the accountability of PLG models, investigating the problem from both model development and deployment perspectives. A holistic framework is proposed for auditing the training data usage of PLG models, detecting neural code generated by PLG models, and determining its attribution to a source model. 
To audit training usage, we introduce a membership inference method based on the likelihood ratio test. We provide several concrete recommendations to enhance the transparency and accountability of PLG models from both technical and regulatory perspectives.
For neural code detection, we propose a learning-based method that can distinguish between human-written code and neural code. Extensive evaluations demonstrate that there exist common fingerprints shared among different PLG models. These findings can help practitioners detect neural code, and mitigate the severe threats recently encountered by the community. 
For neural code attribution, we empirically and theoretically demonstrate the impossibility of reliably solving this one-to-one attribution problem. These findings shed light on the challenges of dissecting neural code accountability. 
To this end, we propose two feasible alternative methods, attribution classification and attribution verification, by relaxing the constraints on either the number of PLG models or the number of neural code snippets. 
Extensive evaluations show the efficacy and robustness of the proposed feasible methods. 
The implementations of the proposed framework are also encapsulated into a new artifact, \textsc{CodeForensic}, to foster further research and inform the technical community at large of the responsible use of generative AI. 

\bibliographystyle{IEEEtranS}
\bibliography{sample-base-short}

\newpage
\section*{Appendix}
% \appendix
\setcounter{section}{0}
\renewcommand{\thesection}{\Alph{section}}

\setcounter{figure}{0}
\renewcommand{\thefigure}{\Alph{section}\arabic{figure}}

\section{Additional Experiment Results}

\subsection{Training Usage Audit}

\begin{figure}[hp]
\centering
\begin{minipage}[]{0.32\textwidth}
\centering
\includegraphics[width=0.96\linewidth]{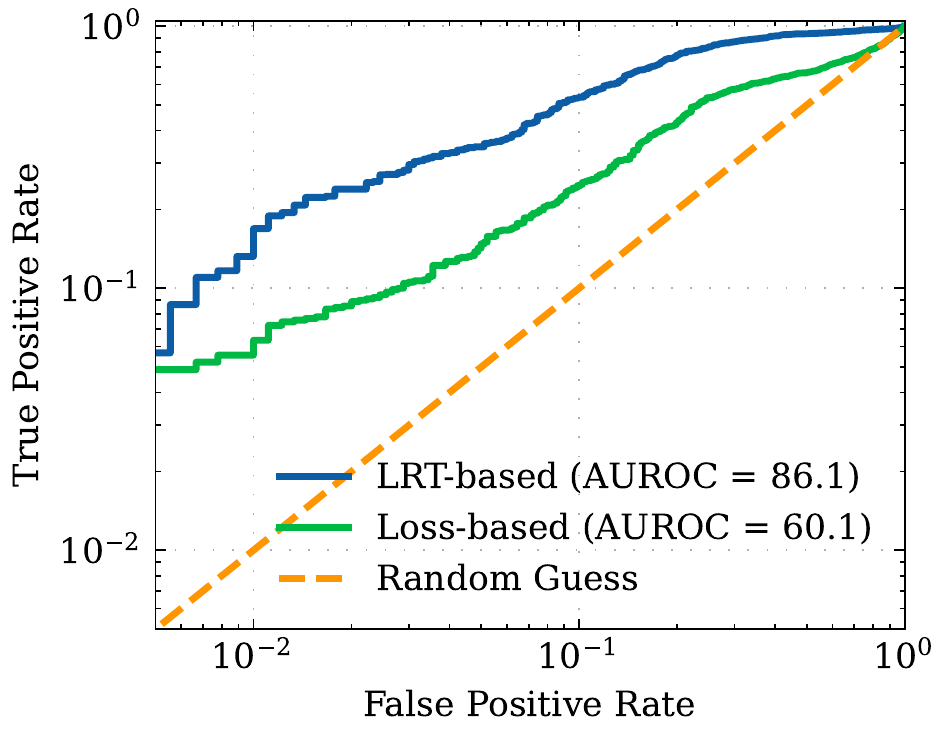}
\end{minipage}
\caption{\label{fig: membership_low_fpr} The ROC curves of the LOSS-based and LRT-based membership inference methods against GPT-Neo on the APPS dataset. }
\vspace{-3mm}
\end{figure}

Following the suggestions of the recent membership inference study~\cite{carlini2022membership}, we also evaluate the performance of different membership inference methods in terms of the True Positive Rate (TPR) at a low False Positive Rate (FPR). As shown in Figure~\ref{fig: membership_low_fpr}, we can observe that the LRT-based method still performs better than the LOSS-based method under tight false positive rate constraints. 

\subsection{Neural Code Detection}

\begin{figure}[htp]
\centering
\subfigure[Temperature]{
\begin{minipage}[]{0.32\textwidth}
\centering
\includegraphics[width=0.96\textwidth]{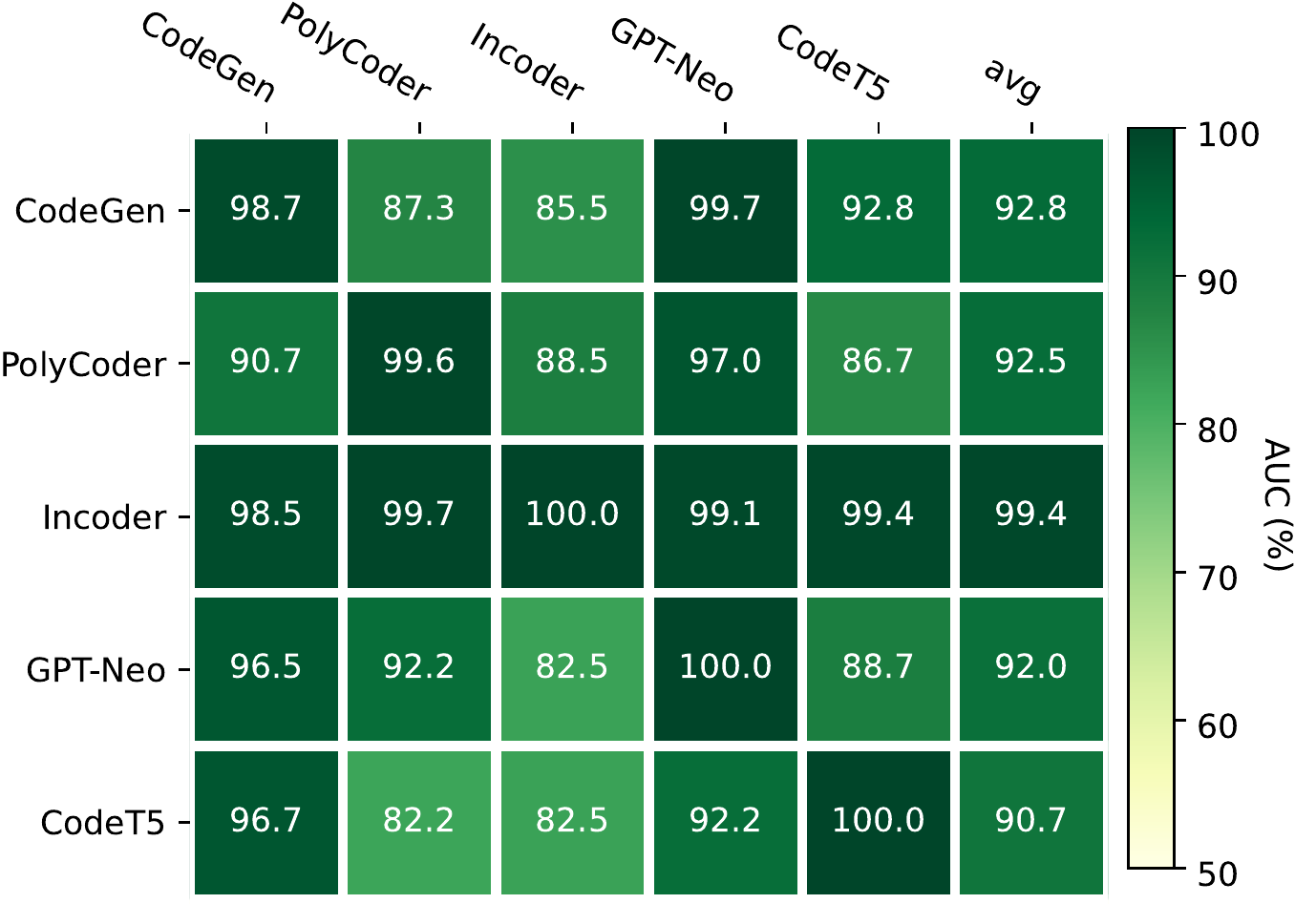}
\label{fig: detection_results_q1_mbpp_temperature}
\end{minipage}}
\subfigure[Nucleus sampling threshold]{
\begin{minipage}[]{0.32\textwidth}
\centering
\includegraphics[width=0.96\textwidth]{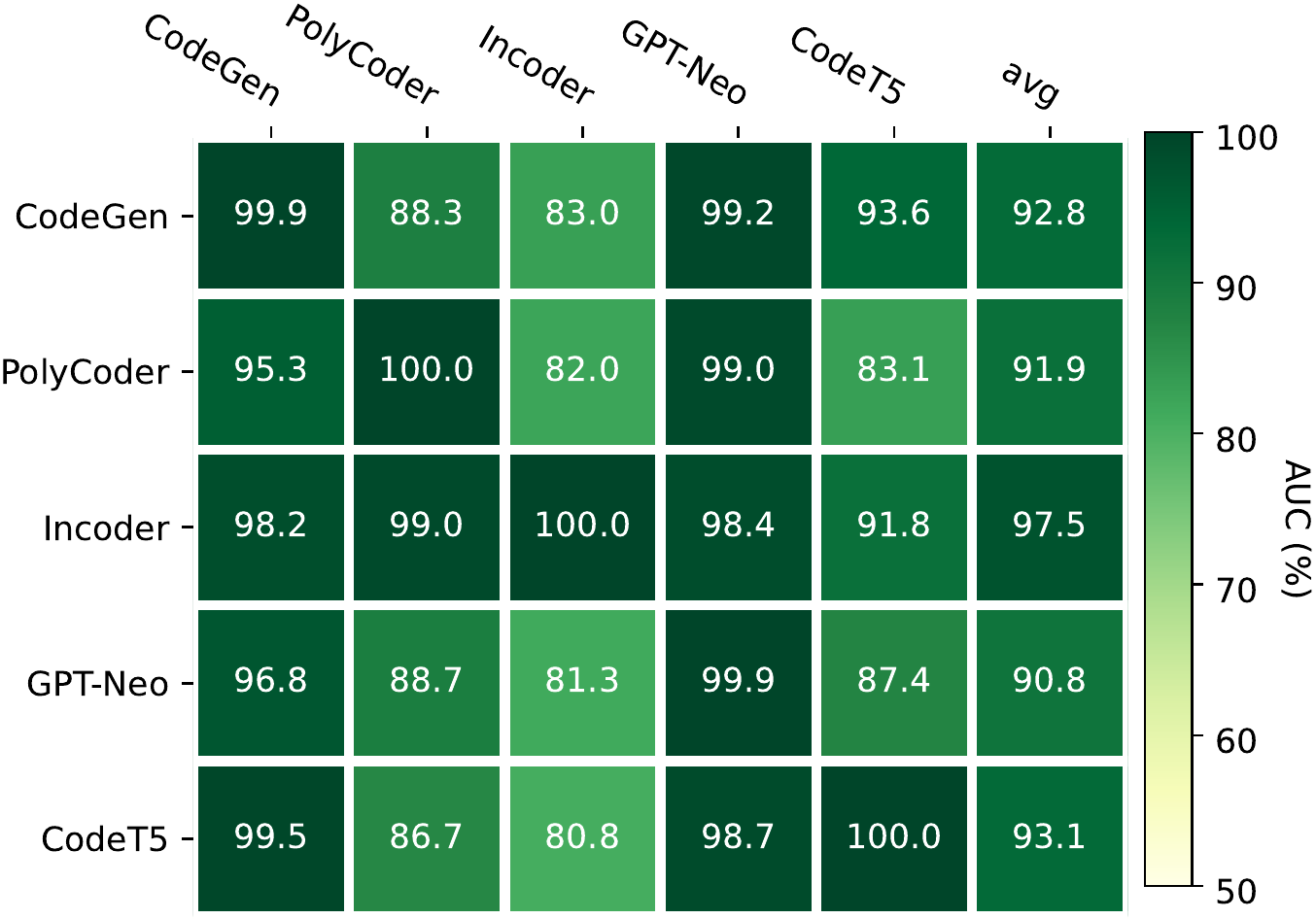}
\label{fig: detection_results_q1_mbpp_top_p}
\end{minipage}}
\vspace{-2mm}
\caption{\label{fig: detection_results_q1_mbpp_ablation} Performance of neural code detection when sampling parameters are different in training and testing stages. (a) Performance of neural code detection when training and testing neural code snippets are generated with temperature $T$=0.2 and $T$=1.0, respectively; (b) Performance of neural code detection on neural code when training and testing neural code snippets are generated with nucleus sampling parameter $p$=0.95 and $p$=1.0, respectively} 
\vspace{-2mm}
\end{figure}

\noindent \textbf{Impact of Sampling Parameters.} As aforementioned, PLG models have optional configuration parameters when generating code snippets. Here, we explore whether the different sampling parameters (\textit{i.e.}, temperature $T$ and nucleus sampling threshold~$p$) affect the detection performance. We conduct two additional experiments: fixing $p=0.95$, we train the detector with $T=0.2$ and test code is generated with $T=1.0$; and fixing $T=0.2$, we train the detector with $p=0.95$ and test code is generated with $p=1.0$. 
We report the results in Figure \ref{fig: detection_results_q1_mbpp_temperature} and Figure~\ref{fig: detection_results_q1_mbpp_top_p}. Compared to the result in Figure 2(b), there is a very slight performance drop (less than 3\% on the average AUC) in Figure \ref{fig: detection_results_q1_mbpp_temperature} when using different sampling temperatures for training and testing code snippets. Similar results are shown in Figure~\ref{fig: detection_results_q1_mbpp_top_p} when varying $p$. Thus, we can conclude that our neural code detector is resilient to differences in sampling parameters.

% \wanlun{shall we omit this part? or I can add an experiment and put it in the appendix.}
We do not evaluate the impact of different decoding strategies (\textit{i.e.}, using top-$k$ sampling rather than nucleus sampling) for two reasons. First, all the PLG models use nucleus sampling as the default decoding strategy, and none of them considers other decoding strategies in performance evaluation~\cite{nijkamp2022codegen, fried2022incoder, xu2022systematic, chen2021evaluating, wang2021codet5}. Second, as is shown in text generation~\cite{ippolito2020automatic, munir2021through}, the detector trained with nucleus sampling demonstrates great transferability and can detect texts generated with other sampling strategies such as top-$k$ sampling with negligible performance decrease.

\subsection{Attribution Classification}

\noindent \textbf{Impact of Training Dataset Size.}
Previous empirical results demonstrate that our attribution classifier achieves excellent performance in identifying the source PLG model of neural code snippets. We now explore the impact of the training data size on attribution performance. Specifically, for Stage 1 in the attribution classification pipeline, we vary the number of generated code snippets ($N$) per model from 4 to 770. Please be noted that the default value of $N$ in the previous evaluation is 770.

\begin{figure}[hp]
\centering
\begin{minipage}[]{0.4\textwidth}
\centering
\includegraphics[width=0.96\linewidth]{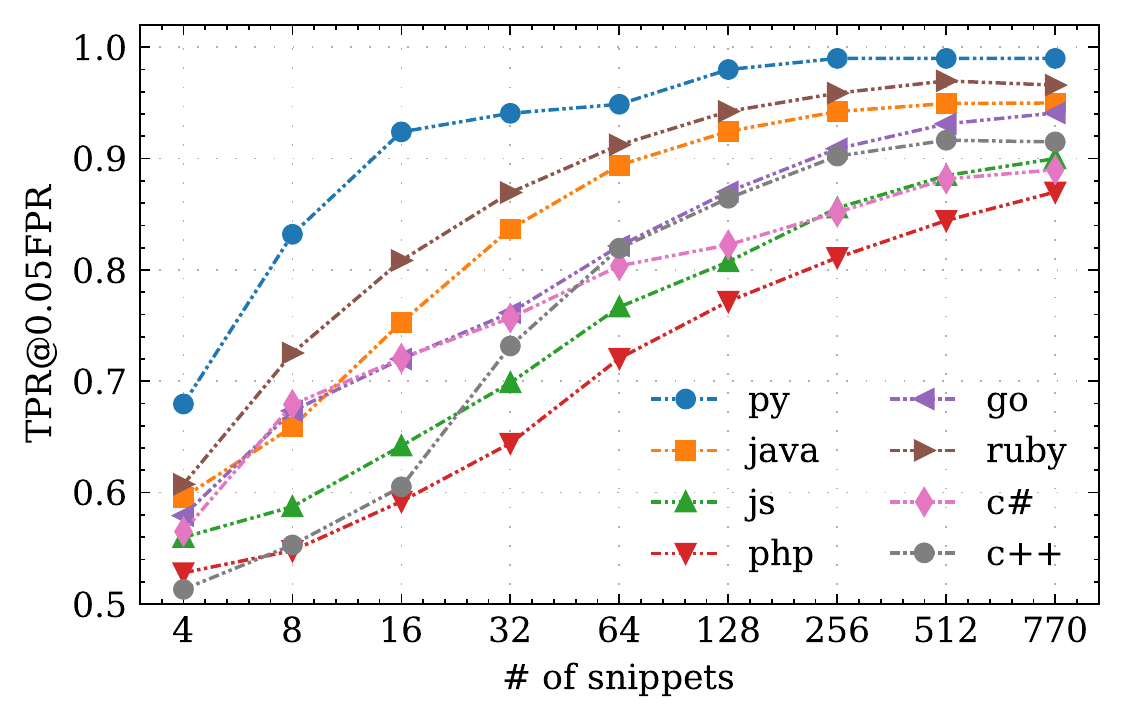}
\end{minipage}
\caption{\label{fig: detection_results_q2_ablation_training_size} Performance of attribution classification under different sizes of training data.}
\vspace{-2mm}
\end{figure}

Figure~\ref{fig: detection_results_q2_ablation_training_size} illustrates the performance of attribution classification under different sizes of training data. We can see that the size of the training data (\textit{i.e.}, generated code snippets from each PLG model)~has a significant effect on the performance of attribution classification in all programming languages. For example, when the training dataset contains only 4 Java code snippets generated from each model, our attribution classifier yields only 59.6\% accuracy, while it can achieve 95.0\% accuracy when the number of available neural code snippets per model increases to 770. In other programming languages, we can observe similar trends that the attribution classification performance becomes better with the increase in the training data size.

Moreover, we can find that our attribution classifier yields $>$70\% accuracy even when the number of generated code snippets per model is relatively small (\textit{i.e.}, around 32$\sim$64 code snippets per model). This is a fairly high performance compared to the five-class random guess accuracy of 20\%. A similar phenomenon is observed in studies of authorship attribution of human-written source code~\cite{abuhamad2018large, bogomolov2021authorship}. For example, Bogomolov \textit{et al.}~\cite{bogomolov2021authorship} shows that the deep-learning-based code authorship attribution method achieves a high accuracy score, outperforming other baselines, even though the median size of Java code snippets per author is only 54.

\subsection{Attribution Verification}

\begin{figure*}[htp]
% \centering
\begin{minipage}[]{0.62\textwidth}
\centering
\subfigure[Temperature]{
\begin{minipage}[]{0.48\linewidth}
\centering
\includegraphics[width=0.96\linewidth]{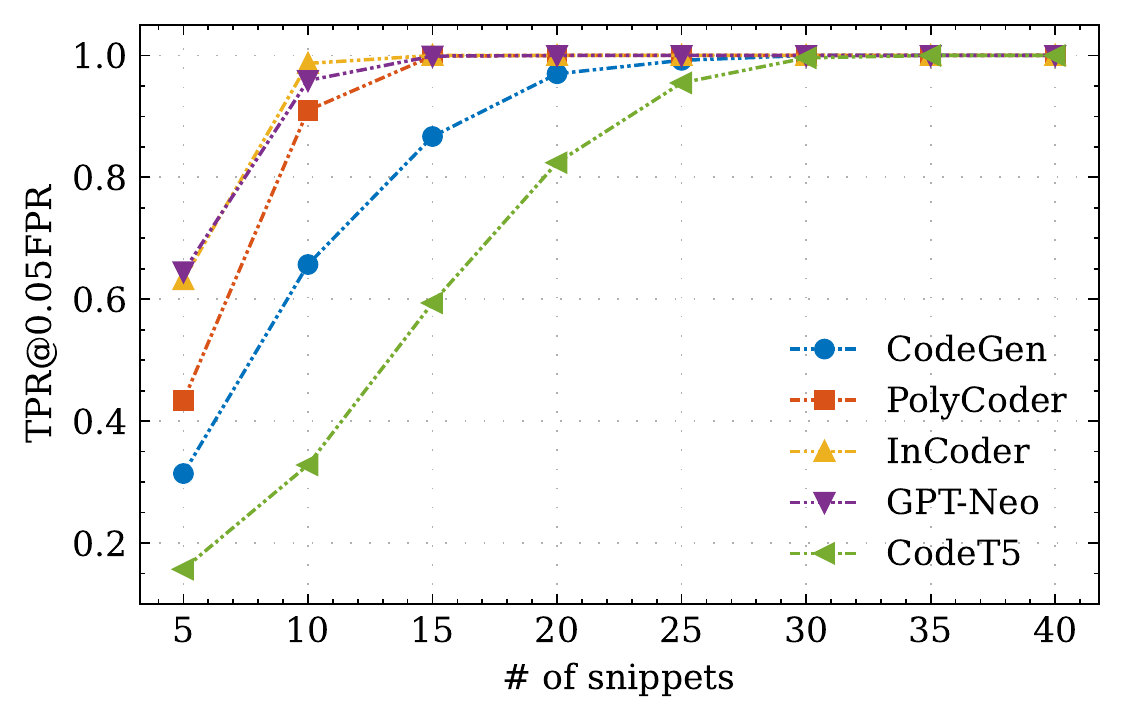}
\label{fig: detection_results_q3_mmd_ablation_temperature}
\end{minipage}}
\subfigure[Nucleus sampling threshold]{
\begin{minipage}[]{0.48\linewidth}
\centering
\includegraphics[width=0.96\linewidth]{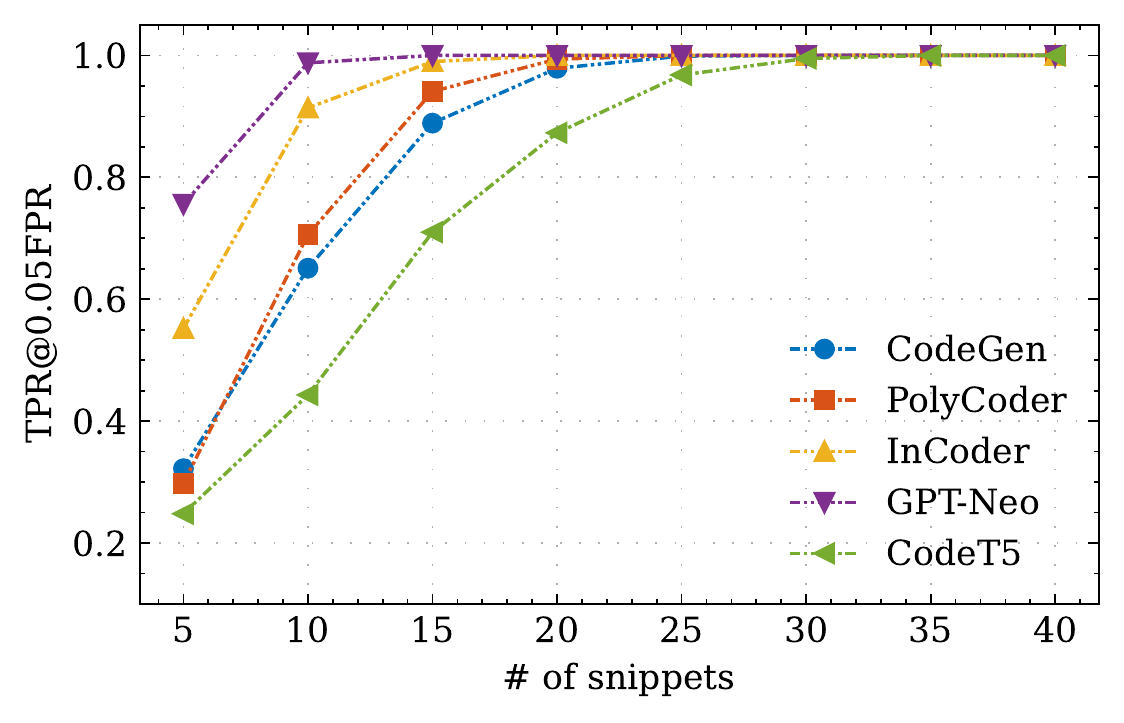}
\label{fig: detection_results_q3_mmd_ablation_top_p}
\end{minipage}}
\vspace{-1mm}
\caption{\label{fig: detection_results_q3_mbpp_ablation} Impact of sampling parameters in attribution verification. (a) code snippets in $\mathcal{X}$ and $\mathcal{X}^{\prime}$ are generated with temperature $T=0.2$ and $T=1.0$, respectively; and (b) code snippets in $\mathcal{X}$ and $\mathcal{X}^{\prime}$ are generated with nucleus sampling parameter $p=0.95$ and $p=1.0$, respectively.} 
\vspace{-2mm}
\end{minipage}
\quad
\begin{minipage}[]{0.30\textwidth}
\centering
\includegraphics[width=0.96\linewidth]{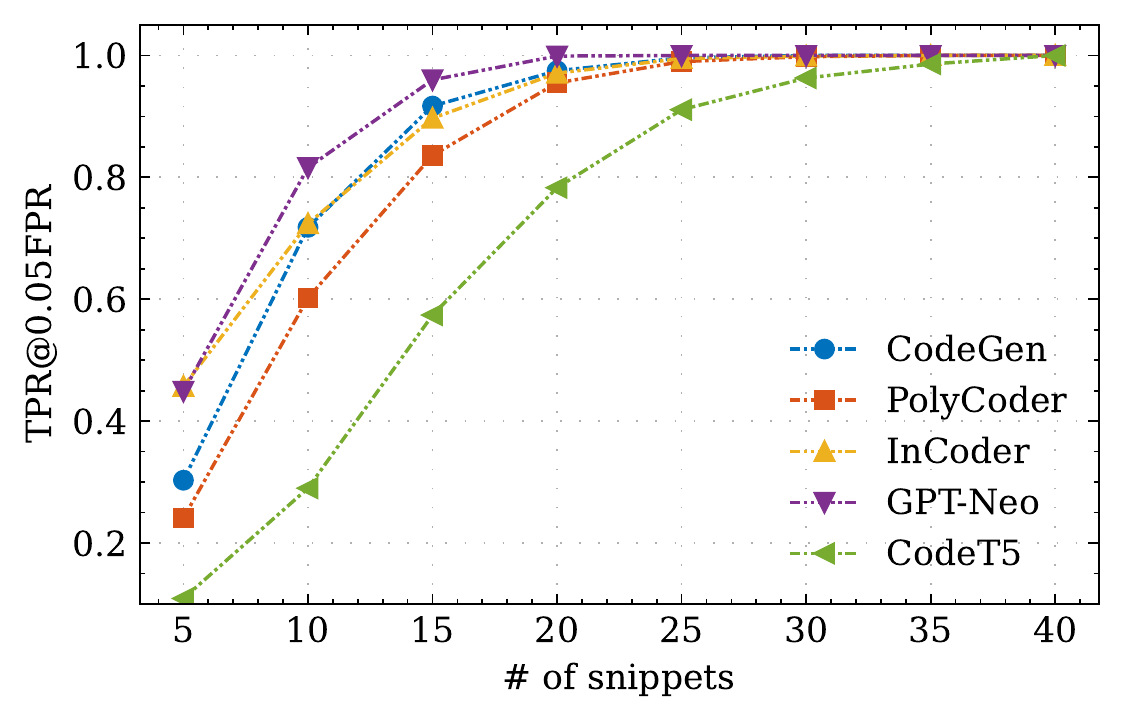}
\vspace{2.5mm}
\caption{\label{fig: detection_results_q3_mmd_ablation_bert} Performance of attribution verification when using BERT as a feature extractor.}
\vspace{-2mm}
\end{minipage}
\end{figure*}

\noindent \textbf{Impact of Sampling Parameters.}
Here, we explore whether different sampling parameters affect the verification performance. 
Similar to the previous evaluation in Section~4.4, we use the default sampling parameters, \textit{i.e.}, $T$=0.2 and $p$=0.95, to generate code snippets for $\mathcal{X}$. When generating testing code snippets for the sample set $\mathcal{X}^{\prime}$, we either change $T$ to 1.0 or $p$ to 1.0 while the other parameter is fixed at the default value. Figure~\ref{fig: detection_results_q3_mmd_ablation_temperature} illustrates the evaluation results using the MBPP dataset when varying the sampling temperature of testing code snippets. Compared to the results in Figure~6(a), we can observe that there is only a very slight performance drop when the sampling temperature is different. For example, TPR drops about 8\% on average when the number of testing code snippets is 5, but the gap narrows to 3\% when the number of testing code snippets increases to 15. A similar trend can be observed in Figure~\ref{fig: detection_results_q3_mmd_ablation_top_p}, where the nucleus sampling threshold $p$ is different for code snippets in $\mathcal{X}$ and $\mathcal{X}^{\prime}$. Therefore, we can conclude that our attribution verification method is resilient to differences in sampling parameters.

\noindent \textbf{Impact of Code Representation Extraction Model.} Our attribution verification relies on a code representation extraction model to transform the raw discrete code data into numerical features. Here, we evaluate the attribution performance with different pre-trained language models, \textit{i.e.}, UniXcoder, and BERT. We report the results in Figure~\ref{fig: detection_results_q3_mmd_ablation_bert}. 
We can obverse that attribution verification using the BERT-based extractor achieve almost the same performance as the UniXcoder-based one when the number of testing code snippets is larger than 20. However, attribution verification using UniXcoder as a code embedding model outperforms the BERT-based one especially when the number of testing code snippets is small, indicating the advantage of utilizing code representations from a programming language-based pre-trained model.

\section{Additional Figures}

\setcounter{figure}{0}
\renewcommand{\thefigure}{\Alph{section}\arabic{figure}} 

\begin{figure}[htp]
\centering
\subfigure[]{
\begin{minipage}[]{0.33\textwidth}
\centering
\includegraphics[width=0.96\textwidth]{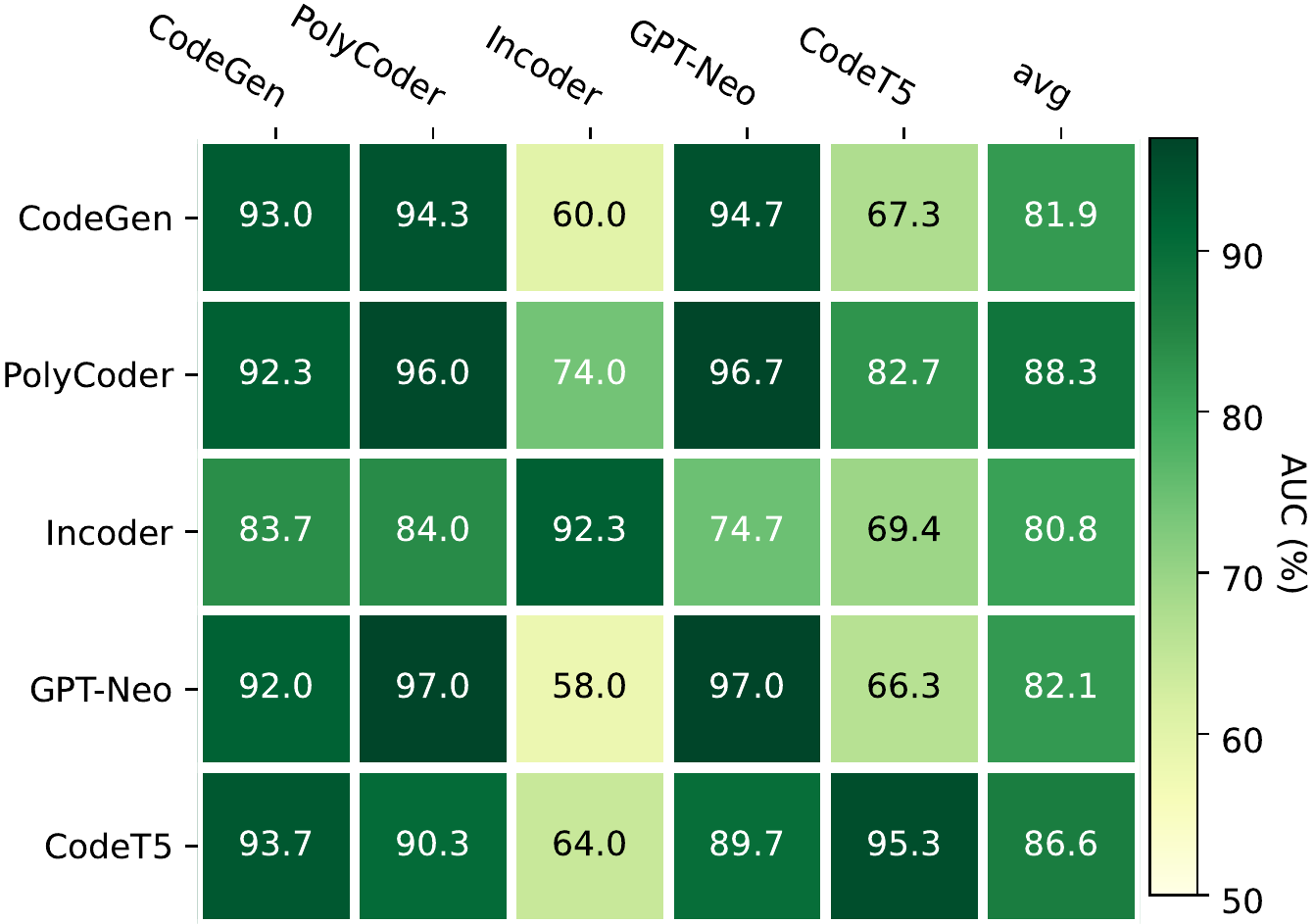}
\end{minipage}}
\subfigure[]{
\begin{minipage}[]{0.33\textwidth}
\centering
\includegraphics[width=0.96\textwidth]{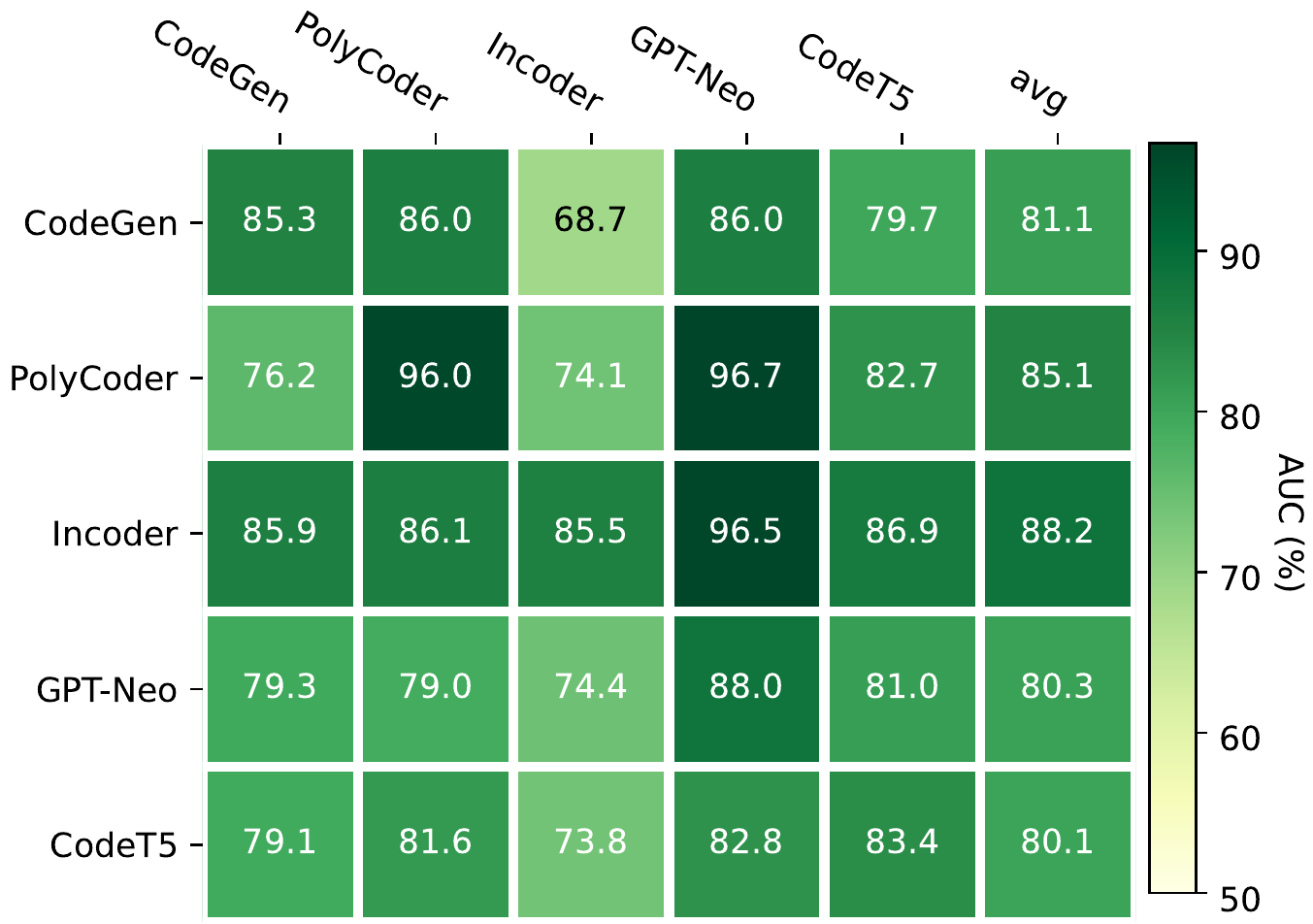}
\end{minipage}}
% \vspace{-4mm}
\caption{\label{fig: detection_results_q1_crossdataset} Performance of neural code detection when training neural code snippets are generated with prompts from MBPP dataset, while testing code snippets are generated with prompts from (a) Humaneval dataset and (b) APPS dataset, respectively;
} 
% \vspace{-3mm}
\end{figure}

\begin{figure}[tp]
\centering
\begin{minipage}[]{0.33\textwidth}
\centering
\includegraphics[width=0.96\textwidth]{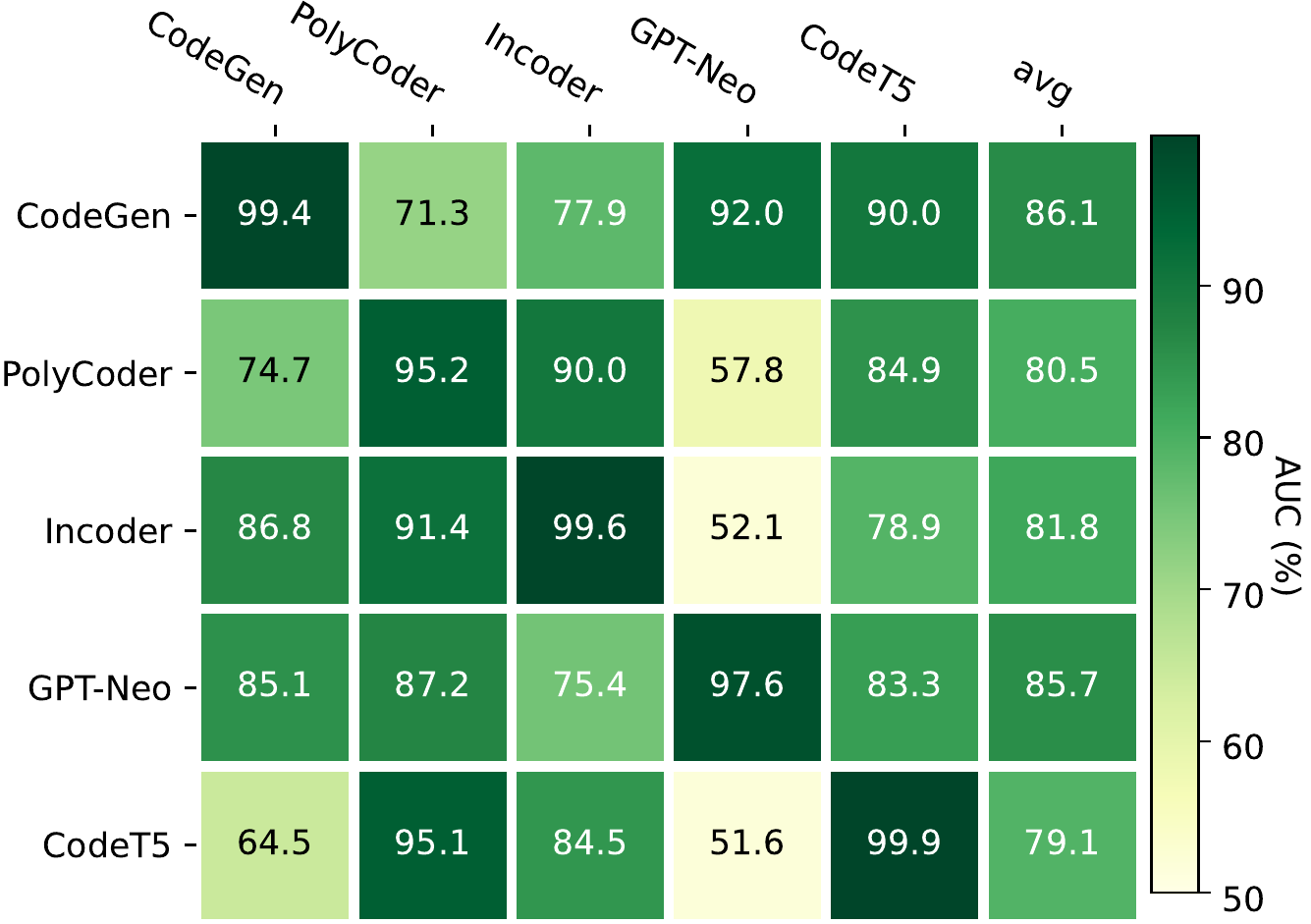}
\end{minipage}
\caption{\label{fig: detection_results_q1_mbpp_bert} Performance of neural code detection using BERT model that is solely pre-trained on natural language corpus.}
\end{figure}

\begin{figure*}[tp]
\centering
\includegraphics[width=0.95\linewidth]{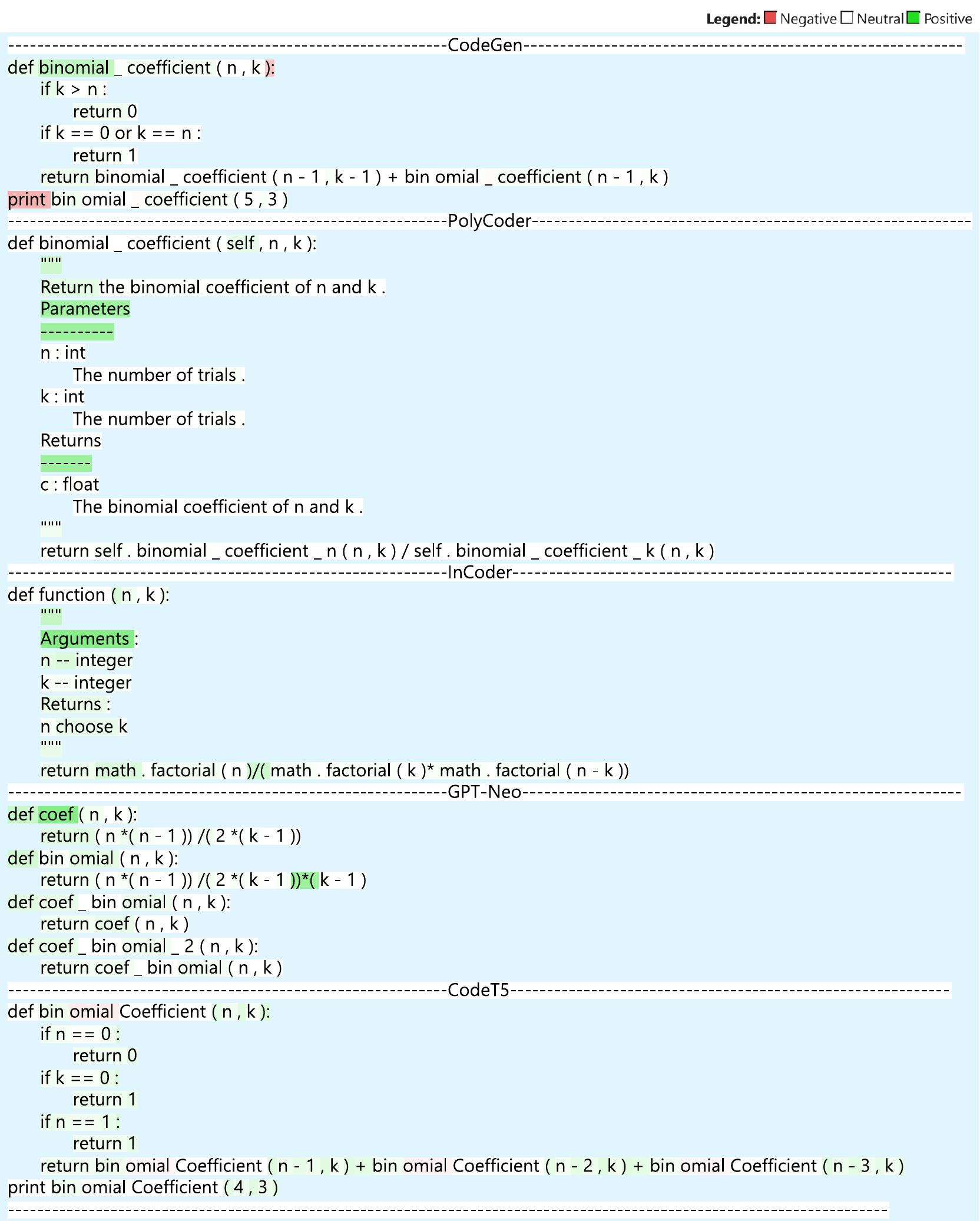}
\caption{Token importance analysis for attribution classification using the model interpretability tool Captum~\cite{kokhlikyan2020captum}. The programming problem (\textit{i.e.}, input prompt for each PLG model) is ``Write a python function to find binomial co-efficient.''}
\label{fig: captum_mbpp_codegen_taskid_28}
\end{figure*}

\end{document}